\documentclass[11pt,a4paper]{article}
\usepackage{jheppub}
\usepackage{amssymb}
\usepackage{amsmath}
\usepackage[normalem]{ulem}

 \usepackage{graphicx}
 \RequirePackage{color}
 \newcommand{\red}[1]{{\color{red}{#1}}}

 \definecolor{MyDarkGreen}{rgb}{0.02,0.60,0.06}

\title{Exact solution of the critical Ising model with special toroidal boundary conditions.}

\author[a]{Armen Poghosyan}
\author[b]{Nickolay Izmailian}
\author[c]{Ralph Kenna}

\affiliation[a]{Yerevan Physics
	Institute, Alikhanian Brothers 2, 375036 Yerevan, Armenia}
\affiliation[b]{Yerevan Physics Institute, Alikhanian Brothers 2, 375036 Yerevan, Armenia}
\affiliation[c]{Applied Mathematics
	Research Centre, Coventry University, Coventry CV1 5FB, UK}

\emailAdd{armenpoghos@yerphi.am} 

\emailAdd{ab5223@coventry.ac.uk; izmail@yerphi.am}

\emailAdd{r.kenna@coventry.ac.uk}

\abstract{
	The Ising model in two dimensions with special toroidal boundary
	conditions is analyzed. These boundary conditions, which we call duality twisted boundary conditions, may be interpreted as inserting a specific defect line (``seam'') in the system, along non-contractible circles of the cylinder, before closing it into a torus. We derive exact expressions for the eigenvalues of a transfer matrix for the critical ferromagnetic Ising model on the M x N square lattice wrapped on the torus with a specific defect line. As a result we have obtained analytically the partition function for the Ising model with such boundary conditions. In the case of infinitely long cylinders of circumference L with duality twisted boundary conditions we obtain the asymptotic expansion of the free energy and the inverse correlation lengths. We find that the ratio of subdominant finite-size correction terms in the asymptotic expansion of the free energy and the inverse correlation lengths should be universal. We verify such universal behavior in the framework of a perturbating conformal approach by calculating the universal structure constant $C_{n1n}$ for descendent states generated by the operator product expansion (OPE) of the primary fields. For such states the calculations of an universal structure constants is a difficult task, since it involves knowledge of the four-point correlation function, which in general is not fixed by conformal invariance except for some particular cases, including the Ising model.}

\arxivnumber{XXXXXXX}

\begin{document}

\maketitle

\section{Introduction}

Finite-size  corrections and scaling  for critical lattice systems, initiated more than four decades ago by Ferdinand, Fisher, and Barber \cite{ferdinand,FerdFisher,barber} have attracted much attention in intervening decades (see Refs. \cite{privman,hu} for reviews).
Finite-size effects are of practical interest due to  recent progress in fine processing technologies,  which has enabled the fabrication of nanoscale materials with novel shapes \cite{nano1,nano2,nano3}. As soon as one has a
finite system one must consider the question of boundary
conditions on the outer surfaces or ``walls'' of the system. As is
well known, the critical behavior near boundaries normally differs
from the bulk behavior.

To understand the effects of boundary conditions, it is especially valuable to study model systems which have exact results (for more details, see \cite{Igloi}).
Two-dimensional models of statistical mechanics have long served
as a proving ground in attempts to understand critical behavior
and to test the general ideas of finite-size scaling. Very few of
them have been solved exactly and the Ising model is the most
prominent example. Since Onsager obtained the exact solution of the two-dimensional
Ising model with cylindrical boundary conditions in 1944
\cite{Onsager}, exact treatments of Ising models on different
two-dimensional surfaces have been continuously attempted to address different two-dimensional topologies \cite{Kaufman,LuWu2001,LuWu1998,Liaw,Brascamp,Brien}.
The partition function for the Ising model on finite lattices has been
calculated exactly for many different boundary conditions,
including toroidal \cite{Kaufman}, the Mobius strip and the Klein
bottle \cite{LuWu2001}, as well as self-dual \cite{LuWu1998}, helical
\cite{Liaw}, and Brascamp and Kunz boundary conditions
\cite{Brascamp}. Additionally systems with cylindrical boundary condition in one direction and free, fixed and mixed boundary condition in another direction have been investigated \cite{Brien}.

The Ising model with a defect line goes back at least to Bariev \cite{Bariev} and McCoy and Perk \cite{Mccoy}. The interpretation in terms of conformal invariance was started by Turban \cite{Turban} and Henkel and Patkos \cite{Henkel1}.
Recently, there has been much progress on understanding conformal
boundary conditions in rational conformal field theories.
This has been caused by its relevance in string and brane theory on the one hand, and in various problems of statistical mechanics and condensed matter on the other. For $sl(2)$ minimal theories, a complete classification has been given
\cite{Behrend,Behrend1} of the conformal boundary conditions on a cylinder.
A complete classification of the conformal boundary condition on the torus has also been given \cite{Affleck1997,Affleck1996,Petkova,Chui}. The key idea is
to compute the partition function on a torus by identifying the
states at the two ends of a cylinder through the trace operation. This may
be interpreted as inserting defect line into the system along non
contractible circles of the cylinder, before closing it into a torus.
The effect of this operation is to ``twist'' the boundary conditions. In statistical mechanics, this is a familiar operation, often referred to as a ``seam''.

In this paper we reported the exact solution of the critical Ising model in two dimensions with a specific defect line. We note that some preliminary results of our investigations have already been announced in a letter \cite{pki}. Our goal is not to discuss defect lines in general, but rather to specialize to a very peculiar type, namely the only one which can be treated \red{in terms of unitary representations of the $c = 1/2$ Virasoro algebra. Note that in general there are an infinite set of boundary conditions associated with defects \cite{Affleck1996,Affleck1997}, which can all be described in terms of unitary representations of the $c = 1$ Virasoro algebra.} The principle of unitarity of the underlying field theory restricts,
through the Kac formula, the possible values of the central charge $c$, and the highest conformal dimensions $\Delta$ and $\bar \Delta$. For the 2D Ising model, we have $c = 1/2$ and the only possible values are $\Delta, \bar \Delta = 0, 1/16, 1/2$.

A central element of the modern theory of bulk critical phenomena is the division into (bulk) universality classes. In general, each bulk universality class of critical phenomena splits into several boundary universality classes. For the Ising universality class there are nine different boundary universality classes. Three of them are associated with the strip geometry and nine boundary universality classes are associated with the cylinder geometry.  

\red{The study of boundary conditions in conformal field theories has attracted much attention in recent decades. The original papers by Cardy and Lewellen \cite{Cardy1991} and Cardy \cite{Cardy1989} on boundary conformal field theory set out the basic properties. For the Ising universality class, which corresponds to a conformal field theory with central charge $c=1/2$, there are three highest weight representations with weights $(0, 1/2, 1/16)$, corresponding to the bulk primary operators ($\mathbb I, \epsilon, \sigma$) representing the unit operator, the energy density and the magnetization, respectively. Thus for the Ising model on a plane we have three different conformally invariant boundary conditions: when the boundary spins are all fixed up, all fixed down or all  are free. They can be denoted by $(+), (-)$ and $(f)$ and they correspond to the conformal dimensions $0, 1/2, 1/16$, respectively. In the strip geometry we have three different pairs of boundary conditions related  by fusion rules to the three possible values of the conformal dimensions $0, 1/16, 1/2$ \cite{Cardy1989}. For example $(--), (-+)$ and $(-f)$ are related to the conformal dimensions $0, 1/2$ and $1/16$, respectively in agreement with Ising fusion rules 
\begin{eqnarray}
\epsilon \otimes \epsilon &=& \mathbb{I} \nonumber\\
\epsilon \otimes \sigma &=& \sigma \nonumber\\
\epsilon \otimes \mathbb{I} &=& \epsilon.
\end{eqnarray}

Thus for the Ising model in the strip geometry we have three different conformally invariant boundary conditions with the conformal dimension $\Delta$ equal to
\begin{eqnarray}
\Delta &=& 0 \hspace{1.3cm} \mbox{for free, fixed $(++)$ or $(--)$ boundary conditions}
\label{eq6.1}\\
\Delta &=& \frac{1}{2} \hspace{1.2cm} \mbox{for fixed $(+-)$ or $(-+)$ boundary conditions}
\label{eq6.3}\\
\Delta &=& \frac{1}{16} \hspace{1cm} \mbox{for mixed boundary conditions.}
\label{eq6.2}
\end{eqnarray}
For fixed $(++)$ (or $(+-)$) boundary conditions the spins are fixed to the same (or opposite) values on two sides of the strip. Mixed boundary conditions correspond to free boundary conditions on one side of the strip, and fixed boundary conditions on the other. In the terminology of surface critical phenomena (for a general review of critical behavior at surfaces, see \cite{Binder}) these three boundary universality classes: free, fixed $(+-)$ and mixed correspond to ``ordinary'', ``extraordinary'' and ``special'' surface critical behavior, respectively.}

\red{Let us now consider the Ising model on a torus defined by modular parameter $\tau=\tau_1+ i\tau_2$. The partition function of the Ising model in CFT can be written as
\begin{equation}
Z_{\rm Ising}={\rm Tr}\, e^{-2\pi\tau_2\left(L_0+\bar L_0-\frac{1}{24}\right)+2\pi i \tau_1\left(L_0-\bar L_0\right)}
\label{ZCFT}
\end{equation}
where $L_0, \bar L_0$ are the Virasoro generators. In terms of the eigenvalues of $L_0$ and $\bar L_0$ we can rewrite the partition function in the form
\begin{equation}
Z_{\rm Ising}=\sum_{\alpha} e^{-2\pi\tau_2\left(\Delta_{\alpha}+\bar \Delta_{\alpha}-\frac{1}{24}\right)+2\pi i \tau_1\left(\Delta_{\alpha}-\bar \Delta_{\alpha}\right)}=\sum_{\alpha} e^{-2\pi\tau_2\left(h_{\alpha}-\frac{1}{24}\right)+2\pi i \tau_1 s_{\alpha}},
\label{ZCFT1}
\end{equation}
where $\Delta_{\alpha},\bar \Delta_{\alpha}$ are the holomorphic and antiholomorphic conformal dimensions of the scaling operators $\phi_{\alpha}$ and $h_{\alpha}=\Delta_{\alpha}+\bar \Delta_{\alpha}$ and $s_{\alpha}=\Delta_{\alpha}-\bar \Delta_{\alpha}$ are known as scaling dimension and conformal spin of $\phi_{\alpha}$ respectively. The scaling operators come in conformal towers built up from the primary operators. If $\Delta, \bar \Delta$ are the conformal dimensions of a primary operator, then the scaling operators in its conformal tower have conformal dimensions of the form $\Delta_n=\Delta+n$ and  $\bar \Delta_m=\bar \Delta+m$, where $n,m \in \mathbb N$. The universal part in finite size corrections of the ground state energy expected from conformal field theory depends on the central charge $c$, the  scaling dimension  $h=\Delta +\bar \Delta$ and conformal spin  $s=\Delta-\bar \Delta$  of the ground state. The scaling dimension  $(h)$ and the conformal spin  $(s)$  of the ground stateare are the universal constants, but may depend on the boundary conditions. In the case of the Ising model ($c=\frac{1}{2}$) the conformal dimensions $\Delta, \bar \Delta$ can take the values $0, \frac{1}{2}$ and $\frac{1}{16}$. Thus for the Ising model on an infinitely long cylinder there are nine possible values of the pair of the conformal dimensions $(\Delta, \bar \Delta)$ and as a consequence nine different boundary conditions. Three of them with conformal spin $s=0$ and the remaining six can be combined into three pairs which can be identified by the absolute value of the conformal spin. Thus for the Ising model on the infinitely long cylinder we have following nine boundary conditions:
\begin{eqnarray}
(\Delta, \bar \Delta) &=& (0,0) \hspace{1.8cm} \mbox{for periodic
boundary conditions}
\label{eq7.1}\\
(\Delta, \bar \Delta) &=& \left(\frac{1}{16}, \frac{1}{16}\right)
\hspace{1cm} \mbox{for antiperiodic boundary conditions}
\label{eq7.2}\\
(\Delta, \bar \Delta) &=& \left( \frac{1}{2}, \frac{1}{2}\right)
\qquad
\label{eq7.3}\\
(\Delta, \bar \Delta)&=& \left(\frac{1}{16}, 0\right) \, \mbox{and} \, \left(0, \frac{1}{16}\right) \hspace{0.5cm}
\mbox{for duality-twisted boundary conditions}
\label{eq7.4}\\
(\Delta, \bar \Delta) &=& \left(\frac{1}{2}, 0\right) \, \mbox{and} \, \left(0, \frac{1}{2}\right) \qquad
\label{eq7.5}\\
(\Delta, \bar \Delta) &=& \left(\frac{1}{2}, \frac{1}{16}\right) \, \mbox{and} \,  \left( \frac{1}{16}, \frac{1}{2}\right)
\qquad \label{eq7.6}
\end{eqnarray}
Note that the pairs of boundary conditions listed in Eqs. (\ref{eq7.4}) - (\ref{eq7.6}) [e.g. $\left(\frac{1}{16}, 0\right)$ and $\left(0, \frac{1}{16}\right)$] can be referred to as complex conjugate boundary conditions following terminology of the Refs. \cite{Grimm,Grimm1993}.}

Past efforts have been focused mainly on periodic and antiperiodic boundary conditions. Little attention has been paid to the boundary condition given by Eqs. (\ref{eq7.3}) - (\ref{eq7.6}). In this paper we will try to fill partially this gap and consider one of these sets of boundary conditions, namely those given by Eq. (\ref{eq7.4}) with $(\Delta, \bar \Delta)= \left(\frac{1}{16}, 0\right)$. We will refer to such boundary conditions as duality-twisted boundary conditions since that name have been given to boundary condition with  $(\Delta, \bar \Delta)= \left(\frac{1}{16}, 0\right)$ for the case of the Ising quantum chain and have been considered in Ref.~\cite{Schutz}. Furthermore, the complete spectrum of the Hamiltonian for finite chains of length $N$ with this set of boundary conditions is obtained exactly \cite{Grimm}. 

Although many theoretical results are now known about the critical
exponents and universal relations among the leading critical
amplitudes, not much information is available on ratios among the
amplitudes in finite-size correction terms \cite{Aharony,Aharony1,Aharony2}. New
universal amplitude ratios for finite-size corrections of the
two-dimensional Ising model on square, triangular and honeycomb
lattices with periodic, antiperiodic, free, fixed and mixed
boundary conditions have been recently presented
\cite{Izmailian2001,Izmailian2009a,Izmailian2009b,Izmailian2012,Izmailian2010}.

In this paper we present a new set of universal amplitude ratios for the finite-size corrections of the two-dimensional Ising model on square lattices for duality twisted boundary conditions and show that such universal behavior is correctly reproduced by the conformal perturbative approach.  We also extend the calculation of the universal structure constants for descendent states generated by the OPE of the primary fields. For such states the calculations of the universal structure constants is a difficult task, since it involves the knowledge of the four-point correlation function, which in general does is fix by conformal invariance except for some particular cases, including the Ising model.

Our objective in this paper is to study the Ising model in two dimensions with particular toroidal boundary conditions, the so-called duality-twisted boundary conditions. The paper is organized as follows. In Sec. \ref{ising}  we introduce the Ising model on the square lattice with duality-twisted boundary conditions.
In Sec. \ref{transfer}  we define the transfer matrix of the model.
In Sec. \ref{eigenvalue} we derive exact expressions for the eigenvalues of the transfer matrix for the critical ferromagnetic Ising model on the $M \times N$ square lattice.
We also derive the exact expression for the conformal partition function in Sec. \ref{partfunc}.
In Sec. \ref{univratio} we consider the limit $N \to \infty$ for which we obtain the expansion of the free energy and inverse correlation lengths for the Ising model on an infinitely long cylinder of circumference $L_x$ with duality twisted boundary conditions.  We find that ratio of subdominant finite-size correction terms should be universal and in
Sec. \ref{pcft} and Sec. \ref{Cnnn} we show that such universal behavior is correctly reproduced by the conformal perturbative
approach. Our main results are summarized and discussed in Sec. \ref{concl}.

\section{Ising model with specific defect line (``seam'')}
\label{ising}

The Ising model is one of the classical models in statistical mechanics.
The anisotropic two-dimensional Ising model on a lattice with periodic boundary conditions and with a specific defect line (``seam'') was formulated in Refs. \cite{Brien,Pearce2001}. For the Ising model the seams are labelled by the Kac labels $(r,s)$.
There are six possible partition functions $Z_{(r,s)}(q)$ labelled by $(r,s)= (1,1), (1,2), (1,3), (2,1), (2,2), (2,3)$  and for  three of them, namely, for the seams $(r,s)= (1,1), (1,2)$ and $(1,3)$, the partition functions $Z_{(r,s)}(q)$ are obtained numerically to very high precision in \cite{Pearce2001}:
\begin{eqnarray}
Z_P=Z_{(1,1)}(q) &=& \left|\chi_0(q)\right|^2+\left|\chi_{\frac{1}{2}}(q)\right|^2+\left|\chi_{\frac{1}{16}}(q)\right|^2 \label{ZP}\\
Z_{(1,2)}(q) &=&  \left[\chi_0(q)+\chi_{\frac{1}{2}}(q)\right]\chi_{\frac{1}{16}}(q)^*+
\left[\chi_0(q)+\chi_{\frac{1}{2}}(q)\right]^*\chi_{\frac{1}{16}}(q) \label{Z12}\\
Z_A=Z_{(1,3)}(q)&=& \chi_0(q)\chi_{\frac{1}{2}}(q)^*+\chi_0(q)^*\chi_{\frac{1}{2}}(q)+\left|\chi_{\frac{1}{16}}(q)\right|^2  \label{ZA}
\end{eqnarray}
where $\chi_0 \equiv \chi_0(q)$, $\chi_{\frac{1}{2}} \equiv \chi_{\frac{1}{2}}(q)$ and $\chi_{\frac{1}{16}} \equiv \chi_{\frac{1}{16}}(q)$ are the chiral characters and $q$ is the modular parameter. The $(r, s) = (1, 1)$ and $(1, 3)$ seams reproduce the well known partition function of the Ising model with  periodic and antiperiodic boundary
conditions respectively.
In contrast, little attention has been paid to the boundary conditions given by the seam $(r, s) = (1, 2)$. In what follows we will give the exact solution for the Ising model with seams $(r,s)=(1,2)$ and show that they correspond to duality-twisted boundary conditions. \red{In our calculations we will follow the method of the Ref. \cite{Pearce2001}.}

\begin{figure}[t]
\begin{center}
\includegraphics[height=4.726cm,width=7cm,angle=0]{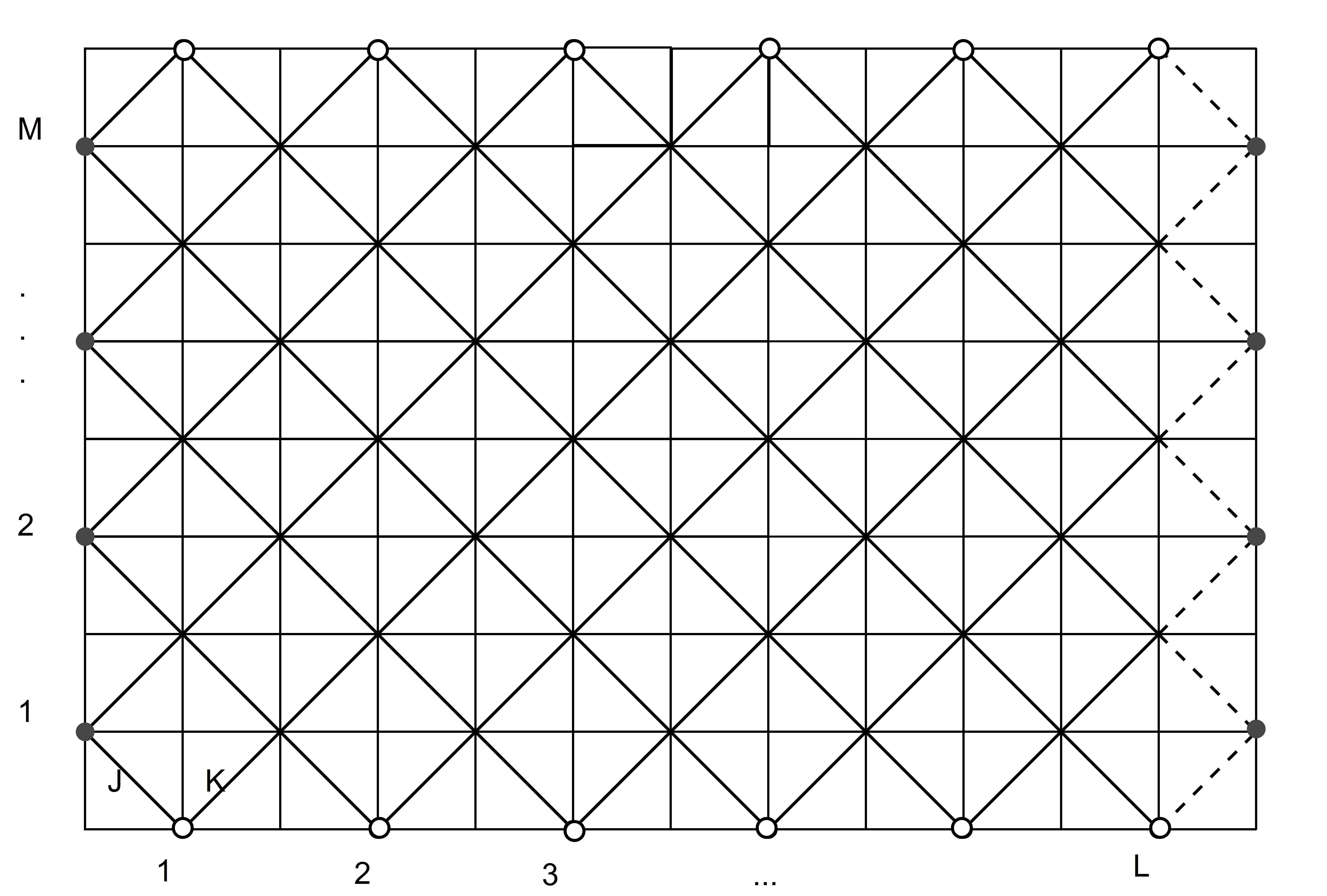}
\end{center}
\caption{
\label{Fig1}
The lattice is constructed with $2L$ regular faces in each row and $2M$ regular faces in each column. The lattice has two sublattices which can be even or odd. The heights are even on the even sublattice and odd on the odd sublattice. On the even sublattice the heights are fixed to the value $2$. On the odd sublattice we identify the state $a=1$ with the usual $+$ Ising state and $a=3$ with the usual $-$ Ising state. The odd sublattice can be considered as the square lattice rotated by $45$ degrees, in which each row has $L$ faces and each column has $M$ faces. The faces on the odd sublattice are twice as large as regular faces. The line defect is inserted along the rightmost columnar edge (dashed line) before closing the lattice into a torus with periodic boundary conditions. Periodic boundary conditions are imposed in both directions.}
\end{figure}
 Let us consider a square lattice rotated by $45$ degrees (Fig. \ref{Fig1}), in which each row has $L$ faces and each column has $M$ faces. We insert one defect line (seam) along the last column. The lattice thus consists of $2L-1$ regular (zigzagging) columnar edges and one defect line and $2M$ regular (zigzagging) row edges.
Periodic boundary conditions are imposed in both directions and in Fig. \ref{Fig1} this is represented as identifying  the light (dark) nodes on the first row (column) with the respective light (dark) nodes on the last row (column). The physical dimensions of the lattice, $L_x$ and $L_y$, are given by
\begin{eqnarray}
L_x&=&\sqrt{2}\left(L-\frac{1}{2}\right)\label{Lx}\\
L_y&=&\sqrt{2}\;M\label{Ly}.
\end{eqnarray}
The finite-size partition function for the Ising model $Z_{L_x L_y}$ can be written as
\begin{equation}
Z_{L_x L_y} = \sum_{s}\exp{\left(J\sum_{\langle ij \rangle} s_i s_j+K\sum_{\langle ij \rangle}s_i s_j\right)}
\nonumber
\label{ZIsing}
\end{equation}
where the first sum within the parenthesis is over NW-SE edges, and the second sum over NE-SW edges and spin variable $s_i$ can take the two values $\pm 1$. Since we restrict ourselves to the critical Ising model, we have $\sinh(2J) \sinh(2K) = 1$. This condition can be conveniently parameterized by introducing a so-called spectral parameter $u$, so that $\sinh(2J) = \cot(2u), \sinh(2K) = \tan(2u)$, with $0<u<\pi/4$. The anisotropy parameter is $\zeta$ related to the spectral parameter $u$ through
\begin{equation}
\xi = \sin 4u.
\end{equation}
For the isotropic system $(K = J)$ we have $u = \pi/8$ and $\zeta = 1$.
Now the finite-size partition function $Z_{L_x L_y}$ can be rewritten in the following form
\begin{equation}
Z_{L_x L_y}(u) = Tr\left[T(u)^M\right] =\sum_{n}e^{-M {\bf E}_n(u)}=\sum_{n}e^{-L_y {\cal E}_n(u)}
\end{equation}
where the sum is over all eigenvalues of a transfer matrix $T(u)$, written as $e^{-{\bf E}_n(u)}$ and
\begin{equation}
{\cal E}_n(u)=\frac{\sqrt{2}}{2} {\bf E}_n(u).
\end{equation}

Consider a row-periodic  transfer matrix $T(u)$. If we write the eigenvalues of this transfer matrix as
\begin{equation}
T_n(u)=\exp{\left(-{\bf E}_n(u)\right)}=\exp{\left(-\sqrt{2}\; {\cal E}_n(u)\right)}, \qquad n=0,1,2,...
\label{TransferM}
\end{equation}
then conformal invariance predicts that the leading finite-size corrections to the energies ${\cal E}_n$ take the form \cite{Chui}
\begin{eqnarray}
 &&{\cal E}_n(u)=L_x f_{\rm{bulk}}\nonumber\\
 &&+\frac{2\pi}{L_x}\left[\left(-\frac{c}{12}+\Delta_n+\bar\Delta_n+k_n+\bar k_n\right)\sin(g u)+i(\Delta_n-\bar\Delta_n+k_n-\bar k_n)\cos(g u)\right]+O\left(\frac{1}{L_x}\right)\nonumber\\
\label{energy}
\end{eqnarray}
where $f_{\rm{bulk}}$ is the bulk free energy, $c$ is the central charge (for Ising model $c=\frac{1}{2}$),
$\Delta_n$ and $\bar\Delta_n$ are the conformal dimensions, $k_n, \bar k_n \in N$, label descendent levels and $g$ is Coxeter number, which for Ising model is $g=4$. Eq. (\ref{energy}) can be rewritten as
\begin{eqnarray}
{\cal E}_n(u)=L_x f_{\rm{bulk}}+\frac{2\pi i}{L_x}\left[\left(-\frac{c}{24}+\Delta_n+k_n\right)e^{-i g u}-\left(-\frac{c}{24}+\bar\Delta_n+\bar k_n\right)e^{i g u}\right]+O\left(\frac{1}{L_x}\right).\nonumber\\
\label{energy1}
\end{eqnarray}
The leading finite-size corrections to the ground state energies $E_0$ take the form
\begin{eqnarray}
{\cal E}_0(u)&=&L_x f_{\rm{bulk}}+\frac{2\pi}{L_x}\left[\left(-\frac{c}{12}+\Delta_0+\bar\Delta_0\right)\sin(g u)+i(\Delta_0-\bar\Delta_0)\cos(g u)\right]+O\left(\frac{1}{L_x}\right)
\nonumber\\
&=&L_x f_{\rm{bulk}}+\frac{2\pi i}{L_x}\left[\left(-\frac{c}{24}+\Delta_0\right)e^{-i g u}-\left(-\frac{c}{24}+\bar\Delta_0\right)e^{i g u}\right]+O\left(\frac{1}{L_x}\right).
\label{energy3}
\end{eqnarray}

\section{The transfer matrix of the model}
\label{transfer}
Unitary $c<1$ CFT models admit a full classification known as the ADE classification \cite{T1,T2,T3}. Lattice realizations of these models have been constructed in \cite{T4}. The ADE Dinkin diagrams play a central role. The Boltzmann weights \cite{Pearce2001} of the models are prescribed to the faces of a regular square lattice
\begin{eqnarray}
\label{BW}
W\left(\begin{array}{cc|}d & c \\ a & b \end{array}\;u\right)
=s_1(-u)\delta_{ac}+
s_0(u)\frac{\sqrt{\psi_a\psi_c}}{\psi_b}.\delta_{bd}
\end{eqnarray}
Here $a,b,c,d$ are the spin states that take values from $G=A,D,E$ graphs, $u$ ($0<u<\lambda$) is the spectral parameter, $\lambda=\frac{\pi}{g}$ is the crossing parameter, and $g$ is the Coxeter number of G. For $s_k(u)$
\begin{eqnarray}
s_k(u)=\frac{\sin(u+k\lambda)}{\sin\lambda},
\end{eqnarray}
$\psi_a$ are the entries of Perron-Frobenius eigenvector (i.e. normalized eigenvector with positive components) of the adjacency matrix $G$.

\begin{figure}[t]
\begin{center}
\includegraphics[height=0.865cm,width=7cm,angle=0]{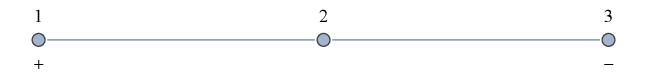}
\caption{
\label{Fig2}
$A_3$ Dynkin diagram}
\end{center}
\end{figure}

\begin{figure}[htb]
\begin{center}
\includegraphics[height=1.43cm,width=7cm,angle=0]{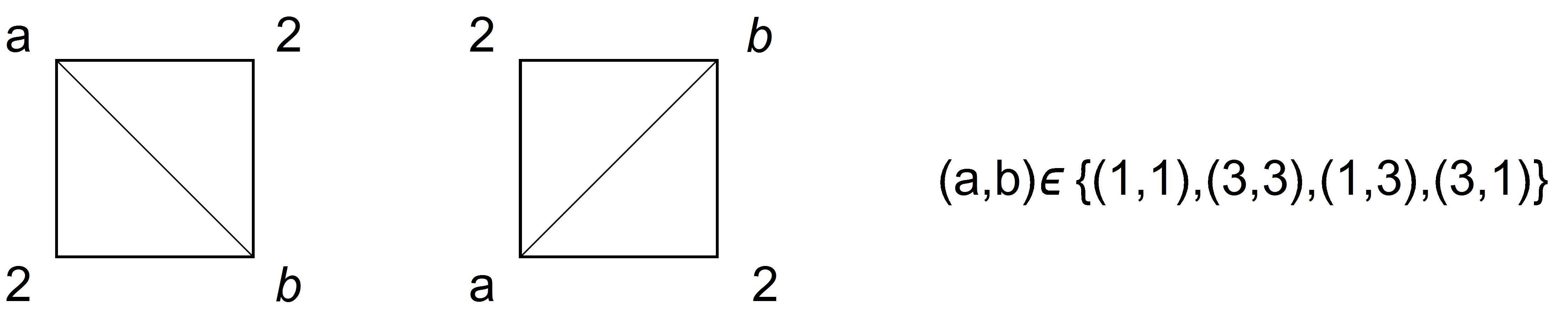}
\caption{
\label{Fig3}
Allowed face configurations}
\end{center}
\end{figure}

\begin{figure}[htb]
\begin{center}
\includegraphics[height=2.383cm,width=7cm,angle=0]{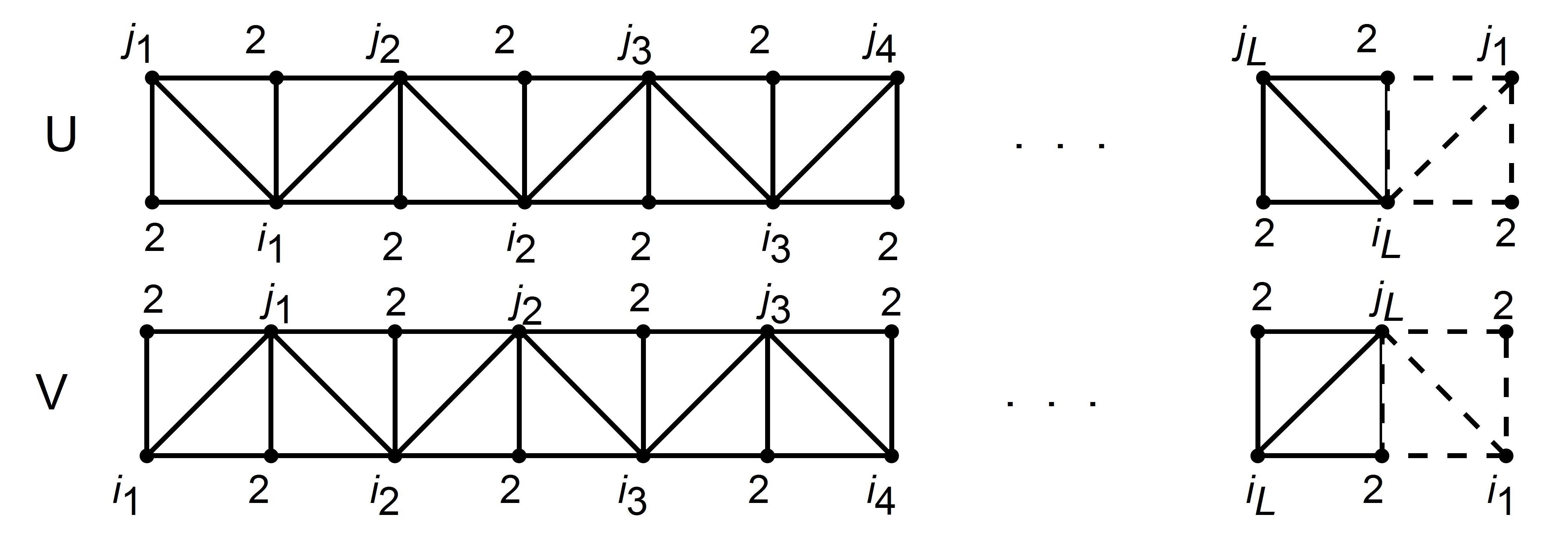}
\end{center}
\caption{
\label{Fig4}
$U$ and $V$ transfer matrices. The rows of the transfer matrices consist of $2L-1$ regular faces followed by single seams. Single seams are represented by dashed lines. The indices $i_{k}$, $j_{k}$ take values $1$ or $3$.}
\end{figure}

\begin{figure}[t]
\begin{center}
\includegraphics[height=1.816cm,width=7cm,angle=0]{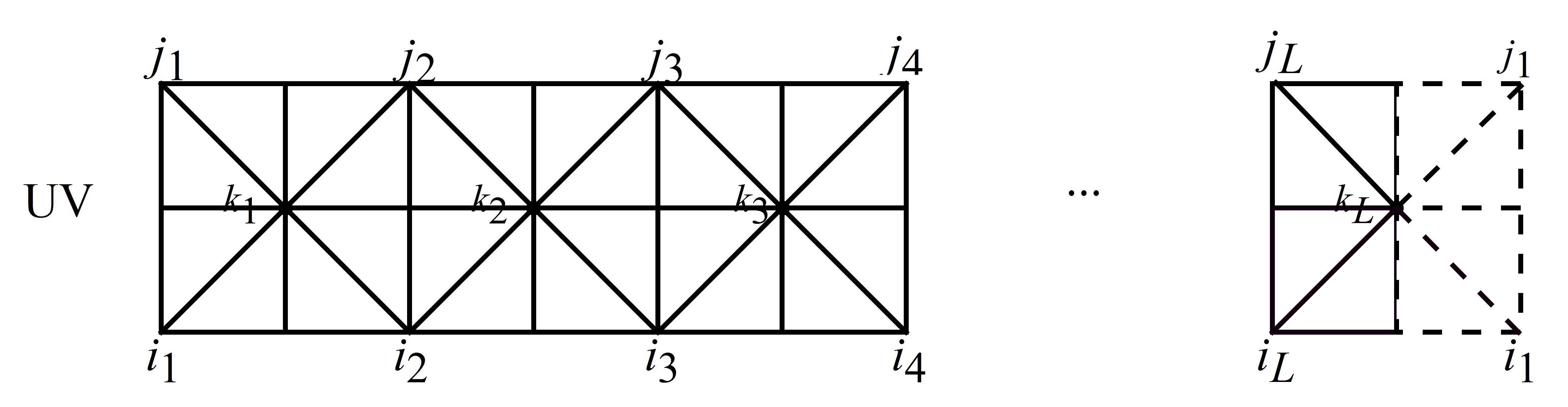}
\end{center}
\caption{
\label{Fig5}
The $UV$ double row transfer matrix consists of $L-1$ blocks and the last dashed block.}
\end{figure}

In this theory the critical Ising model corresponds to $\mathcal{M}(A_2,A_3)$.
So the Ising model is related to the Dynkin diagram $A_{3}$ (see Fig. \ref{Fig2})
whose Coxeter number $g=4$. The $A_{3}$ adjacency matrix is
\begin{eqnarray}
G=\begin{pmatrix} 0 & 1& 0 \\  1 & 0& 1\\  0 & 1& 0  \end{pmatrix} .
\end{eqnarray}
Its Perron-Frobenius eigenvector is
\begin{eqnarray}
\psi=\left(
\begin{array}{c}
 \frac{1}{2} \\ \frac{\sqrt{2}}{2}  \\  \frac{1}{2}
\end{array}
\right).
\end{eqnarray}
The $A_3$ model is one of the Andrews-Baxter-Forrester models \cite{Andrews} $A_L$ with $L=3$. In the $A_3$ model, the spins $a,b,c,d,...$ assigned to the sites of the lattice take heights from the set ${1,2,3}$ (from $A_3$ Dynkin diagram) and satisfy the adjacency condition that heights on adjacent sites must differ by $\pm 1$. In the $A_3$ model, the square lattice has two sublattices which can be even or odd. The heights are even on the even sublattice and odd on the odd sublattice. On the even sublattice the heights are fixed to the value 2. On the odd sublattice we identify the state $a=1$ with the usual $+$ Ising state and $a=3$ with the usual $-$ Ising state. From the $A_{3}$ Dynkin diagram (Fig. \ref{Fig2}) we get the allowed face configurations (see Fig. \ref{Fig3}).
We can calculate the weights of these faces by inserting the
respective parameters in equation (\ref{BW}). Thus we get
 \begin{eqnarray}
 \label{W1}
W\left(\begin{array}{cc|}  1 & 2 \\  2 & 1 \end{array}\;u\right)=
W\left(\begin{array}{cc|}  3 & 2 \\  2 & 3 \end{array}\;u\right)=
\sqrt{2}\sin{\left(u+\frac{\pi}{4}\right)}\nonumber\\
W\left(\begin{array}{cc|}  3 & 2 \\  2 & 1 \end{array}\;u\right)=
W\left(\begin{array}{cc|}  1 & 2 \\  2 & 3 \end{array}\;u\right)=
\sqrt{2}\cos{\left(u+\frac{\pi}{4}\right)}
\end{eqnarray}
\begin{eqnarray}
\label{W2}
W\left(\begin{array}{cc|}  2 & 1 \\  1 & 2 \end{array}\;u\right)=
W\left(\begin{array}{cc|}  2 & 3 \\  3 & 2 \end{array}\;u\right)=
\cos{\left(u\right)}\nonumber\\
W\left(\begin{array}{cc|}  2 & 3 \\  1 & 2 \end{array}\;u\right)=
W\left(\begin{array}{cc|}  2 & 1 \\  3 & 2 \end{array}\;u\right)=
\sin{\left(u\right)}.
\end{eqnarray}
Obviously one can construct two kinds of transfer matrix (see Fig. \ref{Fig4}).
Here the rows consist of $2L-1$ usual faces followed by single seams (depicted in red). The indices $i_{k}$, $j_{k}$ take values $1$ or $3$. They are ($2^{L}\times2^{L}$) square matrices.
Our boundary conditions are given with seams $(r,s)=(1,2)$, which for the $A_3$ model (Ising model) have the following weights \cite{Pearce2001}
\begin{eqnarray}
\label{BW1}
W^{(1,2)}\left(\begin{array}{cc}d & c \\ a & b \end{array}\right)
=ie^{i\frac{\pi}{8}}\delta_{ac}-
ie^{-i\frac{\pi}{8}}\frac{\sqrt{\psi_a\psi_c}}{\psi_b}\delta_{bd}.
\end{eqnarray}
Thus we get
\begin{eqnarray}
\label{WS1}
W^{(1,2)}\begin{pmatrix}  1 & 2 \\  2 & 1 \end{pmatrix}=
W^{(1,2)}\begin{pmatrix}  3 & 2 \\  2 & 3 \end{pmatrix}&=&
ie^{i\frac{\pi}{8}}-\sqrt{2}ie^{-i\frac{\pi}{8}}\nonumber\\
W^{(1,2)}\begin{pmatrix}  3 & 2 \\  2 & 1 \end{pmatrix}=
W^{(1,2)}\begin{pmatrix}  1 & 2 \\  2 & 3 \end{pmatrix}&=&
ie^{i\frac{\pi}{8}}
\end{eqnarray}
\begin{eqnarray}
\label{WS2}
W^{(1,2)}\begin{pmatrix}  2 & 1 \\  1 & 2 \end{pmatrix}=
W^{(1,2)}\begin{pmatrix}  2 & 3 \\  3 & 2 \end{pmatrix}&=&
ie^{i\frac{\pi}{8}}-\frac{ie^{-i\frac{\pi}{8}}}{\sqrt{2}}\nonumber\\
W^{(1,2)}\begin{pmatrix}  2 & 3 \\  1 & 2 \end{pmatrix}=
W^{(1,2)}\begin{pmatrix}  2 & 1 \\  3 & 2 \end{pmatrix}&=&
-\frac{ie^{-i\frac{\pi}{8}}}{\sqrt{2}}
\end{eqnarray}
Note that the (1,2) seam weights are complex and since the defect line can be interpreted in terms of modified bonds $J'$ and $K'$ it means that bonds $J'$ and $K'$ may be complex. In terms of these quantities we can construct the double row transfer matrix $U(u)V(v)$
(see Fig. \ref{Fig5}).
Where $i_a$, $j_a$, $k_a$ take values ${1,3}$ ($a= 1,...L$), and over the variables $k_1$, $k_2$,$...$, $k_L$  a summation is performed. On other vertexes assume non-fluctuating heights equal to $2$. The double row transfer matrix consists of $L-1$ blue blocks and the last blue-red block which we denote by $G(u,v)$ and $S(u,v)$ respectively. In terms of elemental Boltzmann weights Eqs. (\ref{W1}), (\ref{W2}), (\ref{WS1}), (\ref{WS2}) we get
\begin{eqnarray}
\label{GS}
G\left(\begin{array}{cc|c}  j_a & j_{a+1}&u \\ i_a & i_{a+1}&v  \end{array}\right)
=\sum_{k_a={1,3}}
W\left(\begin{array}{cc|}  j_a & 2 \\  2 & k_a \end{array}\;u\right)
W\left(\begin{array}{cc|}  2 & j_{a+1} \\  k_a & 2 \end{array}\;u\right)
W\left(\begin{array}{cc|}  2 & k_a \\  i_a & 2 \end{array}\;v\right)
W\left(\begin{array}{cc|}  k_a & 2 \\  2 & i_{a+1} \end{array}\;v\right)\nonumber\\
S\left(\begin{array}{cc|c}  j_L & j_1&u \\ i_L & i_1&v  \end{array}\right)
=
\sum_{k_L={1,3}}
W\left(\begin{array}{cc|}  j_L & 2 \\  2 & k_L \end{array}\;u\right)
W^{(1,2)}\begin{pmatrix}  2 & j_1 \\ k_L & 2 \end{pmatrix}
W\left(\begin{array}{cc|}  2 & k_L \\  i_L & 2 \end{array}\;v\right)
W^{(1,2)}\begin{pmatrix}  k_L & 2 \\  2 & i_1 \end{pmatrix}.\nonumber\\
\end{eqnarray}
For the matrix elements of the transfer matrix $T(u,v)=U(u)V(v)$ we have
\begin{eqnarray}
\label{TTT}
T^{j_1\;j_2\;j_3\cdots j_L }_{{i_1\;i_2\;i_3\cdots i_L }}(u,v)=
G\left(\begin{array}{cc|c}  j_1 & j_2&u \\ i_1 & i_2&v  \end{array}\right)
G\left(\begin{array}{cc|c}  j_2 & j_3&u \\ i_2 & i_3&v  \end{array}\right)...
S\left(\begin{array}{cc|c}  j_L & j_1&u \\ i_L & i_1&v  \end{array}\right)
\end{eqnarray}
In what follows we will need the weights  $G(u,u+\frac{\pi}{4})$ and $S(u,u+\frac{\pi}{4})$. An elementary calculation ensures that
\begin{eqnarray}
\label{GS2}
&&G\left(\begin{array}{cc}  1 & 1 \\ 1 & 1 \end{array}\right)=
G\left(\begin{array}{cc}  1 & 3 \\ 1 & 3 \end{array}\right)=
G\left(\begin{array}{cc}  3 & 1 \\ 3 & 1 \end{array}\right)=
G\left(\begin{array}{cc}  3 & 3 \\ 3 & 3 \end{array}\right)=
\cos^2(2u)\nonumber\\
&&G\left(\begin{array}{cc}  1 & 1 \\ 3 & 3 \end{array}\right)=
G\left(\begin{array}{cc}  3 & 3 \\ 1 & 1 \end{array}\right)=
-\sin^2(2u)\nonumber\\
&&
G\left(\begin{array}{cc}  1 & 3 \\ 3 & 1 \end{array}\right)=
G\left(\begin{array}{cc}  3 & 1 \\ 1 & 3 \end{array}\right)=
\sin^2(2u)\nonumber\\
&&G\left(\begin{array}{cc}  1 & 1 \\ 1 & 3 \end{array}\right)=
G\left(\begin{array}{cc}  1 & 3 \\ 1 & 1 \end{array}\right)=
G\left(\begin{array}{cc}  3 & 1 \\ 3 & 3 \end{array}\right)=
G\left(\begin{array}{cc}  3 & 3 \\ 3 & 1 \end{array}\right)=
0\nonumber\\
&&G\left(\begin{array}{cc}  1 & 1 \\ 3 & 1 \end{array}\right)=
G\left(\begin{array}{cc}  3 & 3 \\ 1 & 3 \end{array}\right)=
\cos(2u)+\sin(2u)\nonumber\\
&&G\left(\begin{array}{cc}  1 & 3 \\ 3 & 3 \end{array}\right)=
G\left(\begin{array}{cc}  3 & 1 \\ 1 & 1 \end{array}\right)=
\cos(2u)-\sin(2u)\nonumber\\
&&S\left(\begin{array}{cc}  1 & 1 \\ 1 & 1 \end{array}\right)=
S\left(\begin{array}{cc}  1 & 3 \\ 1 & 3 \end{array}\right)=
S\left(\begin{array}{cc}  3 & 1 \\ 3 & 1 \end{array}\right)=
S\left(\begin{array}{cc}  3 & 3 \\ 3 & 3 \end{array}\right)=
\cos(2u)\nonumber\\
&&S\left(\begin{array}{cc}  1 & 1 \\ 3 & 3 \end{array}\right)=
S\left(\begin{array}{cc}  3 & 3 \\ 1 & 1 \end{array}\right)=
-i\sin(2u)\nonumber\\
&&S\left(\begin{array}{cc}  1 & 3 \\ 3 & 1 \end{array}\right)=
S\left(\begin{array}{cc}  3 & 1 \\ 1 & 3 \end{array}\right)=
i\sin(2u)\nonumber\\
&&S\left(\begin{array}{cc}  1 & 1 \\ 1 & 3 \end{array}\right)=
S\left(\begin{array}{cc}  1 & 3 \\ 1 & 1 \end{array}\right)=
S\left(\begin{array}{cc}  3 & 1 \\ 3 & 3 \end{array}\right)=
S\left(\begin{array}{cc}  3 & 3 \\ 3 & 1 \end{array}\right)=
0\nonumber\\
&&S\left(\begin{array}{cc}  1 & 1 \\ 3 & 1 \end{array}\right)=
S\left(\begin{array}{cc}  3 & 3 \\ 1 & 3 \end{array}\right)=
S\left(\begin{array}{cc}  1 & 3 \\ 3 & 3 \end{array}\right)=
S\left(\begin{array}{cc}  3 & 1 \\ 1 & 1 \end{array}\right)=
1.
\end{eqnarray}
It follows from Eq. (\ref{GS2}) that at the special point $v=u+\frac{\pi}{4}$
the only nonzero elements of the transfer matrix are
\begin{eqnarray}
\label{TTTTT}
&&T^{j_1\;j_2\;j_3\cdots j_L }_{{j_1\;j_2\;j_3\cdots j_L }}\left(u,u+\frac{\pi}{4}\right)
=\cos\left(2u\right)^{2L-1}\nonumber\\
&&T^{j_1\;j_2\;j_3\cdots j_L }_{{\tilde {j}_1\;\tilde {j}_2\;\tilde {j}_3\cdots \tilde {j}_L }}\left(u,u+\frac{\pi}{4}\right)
=i\left(-1\right)^{L}\sin\left(2u\right)^{2L-1}
\end{eqnarray}
where $\tilde {j}_a=3$ if $j_a=1$ and $\tilde {j}_a=1$ if $j_a=3$.

Thus the double row transfer matrix $U(u)V(v)$ satisfies the functional equation
\begin{eqnarray}
\label{UV}
U(u)V\left(u+\frac{\pi}{4}\right)=\cos\left(2u\right)^{2L-1}I+i\left(-1\right)^{L}\sin
\left(2u\right)^{2L-1}R,
\end{eqnarray}
where  $I$ is the Identity matrix, and R is a square matrix with anti-diagonal
entries equal to 1 with remaining entries $0$ (both are ($2^{L}\times 2^{L}$) matrices).

\section{Calculation of the Eigenvalues}
\label{eigenvalue}
Let us introduce the  matrix $P(u)$
\begin{eqnarray}
\label{P}
P(u)= \begin{pmatrix} 0 & V(u)  \\ U(u) & 0 \end{pmatrix} .
\end{eqnarray}
It is easy to check that $U^{\dagger}(u)U(u)=V(u)V^{\dagger}(u)$ and,
 $U(u)U^{\dagger}(u)=V^{\dagger}(u)V(u)$ from which we deduce that the Matrix
 $P(u)$ is normal i.e. $P(u)P^{\dagger}(u)=P^{\dagger}(u)P(u)$ so that $P(u)$ is
diagonalizable. Let $\Lambda(u)$ be the eigenvalue and
$\begin{pmatrix}  x_1   \\  x_2   \end{pmatrix}$ the eigenvector of $P(u)$;
\begin{eqnarray}
\label{P-ee}
P(u)\begin{pmatrix}  x_1   \\  x_2   \end{pmatrix} = \begin{pmatrix} 0 & V(u)  \\ U(u) & 0 \end{pmatrix}\begin{pmatrix}  x_1   \\  x_2   \end{pmatrix} =\begin{pmatrix}  V(u)x_2   \\  U(u)x_1   \end{pmatrix}=\Lambda(u)\begin{pmatrix}  x_1   \\  x_2   \end{pmatrix} .
\end{eqnarray}
Evidently
\begin{eqnarray}
\label{UV-ee}
U(u)V(v)x_2=\Lambda(v) U(u)x_1=\Lambda(u)\Lambda(v)x_2\nonumber\\
V(v)U(u)x_1=\Lambda(u) V(v)x_2=\Lambda(u)\Lambda(v)x_1 .
\end{eqnarray}

Since $R^2=1$, the eigenvalues of $R$
are $r=\pm1$.
By acting both sides of the Eq. (\ref{UV}) on the vector $x_2$
we get
\begin{eqnarray}
\label{lambda1}
\Lambda\left(u\right)\Lambda\left(u+\frac{\pi}{4}\right)
&=&\left(\cos2u\right)^{2L-1}+ ir\left(-1\right)^{L}\left(\sin
2u\right)^{2L-1}\nonumber\\
&=&\left(\cos2u\right)^{2L-1}-r\left(i\sin
2u\right)^{2L-1}\qquad \nonumber\\
&=&\left(\frac{e^{2iu}+e^{-2iu}}{2}\right)^{2L-1}
-r\left(\frac{e^{2iu}-e^{-2iu}}{2}\right)^{2L-1} .
\end{eqnarray}
Let us examine the case $r=-1$ (to obtain the eigenvalues
corresponding to $r=+1$, one should simply take the complex conjugates
of the eigenvalues of the case $r=-1$). Changing variable from $u$ to $x  \equiv e^{2iu}$ and noting that if $u\rightarrow u+\frac{\pi}{4}$ then $x \rightarrow ix$ we can obtain from Eq. (\ref{lambda1}) that
\begin{eqnarray}
\label{I1}
\Lambda\left(x\right)\Lambda\left(i x\right)
&=&
(2x)^{-(2L-1)}\left[(x^2+1)^{2L-1}+(x^2-1)^{2L-1}\right]
\end{eqnarray}
To factorize the r.h.s. of Eq. (\ref{I1}) we need to find the $(2L-1)$ solutions of
\begin{eqnarray}
t^{2L-1}+(t-2)^{2L-1}=0 \label{IR} ,
\end{eqnarray}
where $t\equiv x^2+1$. The solutions  $t_k$ for $k=1,2,...,2L-1$  can be expressed through the $2L-1$ roots of $(-1)$  and are
\begin{eqnarray}
&&t_k=-\frac{2 e^{\frac {i\pi (2k-1)}{2L-1}}}{1-e^{\frac{i\pi (2k-1)}{2L-1}}}.
\end{eqnarray}
Then we obtain
\begin{eqnarray}
\label{IR1}
&&t^{2L-1}+(t-2)^{2L-1}=\prod_{k=1}^{2L-1}(t-t_k)\nonumber
\end{eqnarray}
which means that
\begin{eqnarray}
\Lambda\left(x\right)\Lambda\left(i x\right)&=&
2^{2(1-L)}x^{-(2L-1)}\prod_{k=1}^{2L-1} \left(x^2-i\cot\left(
\frac{\pi (2k-1)}{2(2L-1)}\right)\right) . \label{Lamb1}
\end{eqnarray}
Splitting this product into two parts,
\begin{equation}
\prod_{k=1}^{2L-1}F(k)=\prod_{k=1}^{L-1}F(k)\prod_{k=L}^{2L-1}F(k) ,
\end{equation}
then shifting the index in the second part as $k = 2 L - k$, the Eq. (\ref{Lamb1}) may
be expressed as
\begin{eqnarray}
\Lambda\left(x\right)\Lambda\left(i x\right)&=&
2^{2(1-L)}x^{-(2L-1)}x^2\prod_{k=1}^{L-1} \left(
x^2-i\cot
\left(
\frac{\pi (2k-1)}{2(2L-1)}
\right)\right)
\left(x^2+i\cot
\left(
\frac{\pi (2k-1)}{2(2L-1)}\right)\right).\nonumber\\ \label{Lamb2}
\end{eqnarray}
Let us replace $x^2 \to -(i x)^2$ in the first factor of the product in Eq. (\ref{Lamb2}). We obtain
\begin{eqnarray}
\Lambda\left(x\right)\Lambda\left(i x\right)&=&\left(2^{-L+1}\right)^{2}(x)^{-L+\frac{3}{2}}(ix)^{-L+\frac{3}{2}}\left(
i^{\frac{3}{2}L-\frac{7}{4}}\right)^{2}\nonumber\\
&& \prod_{k=1}^{L-1}
\left( x^{2}+i\cot \frac{\pi(2k-1)}{2(2L-1)}\right)
\left( (ix)^{2}+i\cot \frac{\pi(2k-1)}{2(2L-1)}\right)\label{Lamb3}.
\end{eqnarray}
Let us now replace $x^2 \to -(i x)^2$ in the second factor of the product in Eq. (\ref{Lamb2}). We obtain another equivalent representation of $\Lambda\left(x\right)\Lambda\left(i x\right)$:
\begin{eqnarray}
\Lambda\left(x\right)\Lambda\left(i x\right)&=&
\left(2^{-L+1}\right)^{2}(x)^{-L+\frac{3}{2}}(ix)^{-L+\frac{3}{2}}\left(
i^{\frac{3}{2}L-\frac{7}{4}}\right)^{2}\nonumber\\
&& \prod_{k=1}^{L-1}
\left( x^{2}-i\cot \frac{\pi(2k-1)}{2(2L-1)}\right)
\left( (ix)^{2}-i\cot \frac{\pi(2k-1)}{2(2L-1)}\right)\label{Lamb4}
\end{eqnarray}
From Eqs. (\ref{Lamb3}) and (\ref{Lamb4}) we can see that $\Lambda\left(x\right)\Lambda\left(i x\right)$ can be written as
\begin{eqnarray}
\Lambda\left(x\right)\Lambda\left(i x\right)&=&
\left(2^{-L+1}\right)^{2}(x)^{-L+\frac{3}{2}}(ix)^{-L+\frac{3}{2}}\left(
i^{\frac{3}{2}L-\frac{7}{4}}\right)^{2} \nonumber\\
&&\prod_{k=1}^{L-1}
\left( x^{2}+i\mu_{k}\cot \frac{\pi(2k-1)}{2(2L-1)}\right)
\left( (ix)^{2}+i\mu_{k}\cot \frac{\pi(2k-1)}{2(2L-1)}\right)\label{LambFinal}
\end{eqnarray}
where $\mu_k=\pm1$ for all $k$. Now, it is easy to see from Eq. (\ref{LambFinal}) that
\begin{eqnarray}
\label{Lambda}
\Lambda(x)=\varepsilon \,2^{-L+1}i^{\frac{3}{2}L-\frac{7}{4}}x^{-L+\frac{3}{2}}
\prod_{k=1}^{L-1}\left( x^{2}+i\mu_{k}\cot \frac{\pi(2k-1)}{2(2L-1)}\right)
\end{eqnarray}
where $\varepsilon=\pm1$. For the eigenvalues of the two row transfer matrices $UV$ of the twisted Ising model we get
\begin{eqnarray}
\label{L2}
\Lambda^2(u)=2^{-2L+2}\:i^{-L+\frac{1}{2}}\:e^{-2i u(2L-3)}\:
\prod_{k=1}^{L-1}\left(e^{i 4u}+i\mu_{k}\cot \frac{\pi(2k-1)}{2(2L-1)}\right)^2
\end{eqnarray}
in which $\mu_k=\pm 1$ can be chosen arbitrarily so we have $2^{L-1}$ eigenvalues. The remaining eigenvalues can be found by taking the complex conjugate of Eq. (\ref{L2}). Altogether we get $2^L$ eigenvalues. Thus we have obtained all $2^L$ eigenvalues of the two row transfer matrices $U(u)V(u)$. Let us denote the eigenvalues of the two row transfer matrices $U(u)V(u)$ by $\lambda(u)$, which is equal to
$$
\lambda(u)=\Lambda^2(u).
$$
Using the identity $\prod_{k=1}^{L-1}\sin^2\frac{\pi(2k-1)}{2(2L-1)}=2^{-2L+2}$  we can get a new form of Eq. (\ref{L2})
\begin{eqnarray}
\lambda=
e^{2 i \left(u-\frac{\pi }{8}\right)} \prod _{k=1}^{L-1} \left(e^{2 i \left(u-\frac{\pi }{8}\right)} \sin \left(\frac{\pi  (2 k-1)}{2 (2 L-1)}\right)+\mu _k e^{-2 i \left(u-\frac{\pi }{8}\right)} \cos \left(\frac{\pi  (2 k-1)}{2 (2 L-1)}\right) \right)^2. \label{l new} 
\end{eqnarray}
Since $\lambda$ is complex quantity, we can represent it as
$$
\lambda=|\lambda|\exp{(i \theta)}.
$$
From Eq. (\ref{l new}) one can easily obtain the following expressions for the absolute value
of the eigenvalue $|\lambda|$ and the argument $\theta$
\begin{eqnarray}
\left|\lambda\right|&=&
\prod _{k=1}^{L-1}
\left[1+ \mu_k \sin (4 u)
\sin \left(\frac{\pi  (2 k-1)}{2 L-1}\right)\right] , \label{abs}\\
\theta&=&2u-\frac{\pi}{4}+\sum_{k=1}^{L-1}\arctan{\frac{\cos(4u)\cos{\left(\frac{\pi  (2 k-1)}{2
L-1}\right)}}{\sin(4u)+\mu_k \sin\left(\frac{\pi  (2 k-1)}{2 L-1}\right)}} .\label{theta}
\end{eqnarray}
Let us now consider the largest eigenvalue $\lambda_0$, which corresponds to the case when all $\mu_k=1$. For the absolute value  of the largest eigenvalue $|\lambda_0|$ we obtain
\begin{eqnarray}
\left|\lambda_0\right|&=&
\prod _{k=1}^{L-1}
\left[1+  \sin (4 u)
\sin \left(\frac{\pi  (2 k-1)}{2 L-1}\right)\right] \label{abs1}\\
&=&
\prod _{k=1}^{L-1}
\left[1+  \sin (4u)\sin \left(\frac{2\pi k}{2 L-1}\right)\right] \nonumber\\
&=&
\prod _{k=0}^{L-1}
\left[1+  \sin (4u)\sin \left(\frac{\pi k}{2 L-1}\right)\right] . \nonumber
\end{eqnarray}
The product $\prod_{k=0}^{2L-2} \left[1+  \sin (4u)\sin \left(\frac{\pi k}{2 L-1}\right)\right]$
can be expressed as
\begin{equation}
\prod_{k=0}^{2L-2} \left[1+  \sin (4u)\sin \left(\frac{\pi k}{2 L-1}\right)\right]=\prod_{k=0}^{L-1}\left[1+  \sin (4u)\sin \left(\frac{\pi k}{2 L-1}\right)\right]^2. \label{transf2}
\end{equation}
Then the absolute value of $\lambda_0$ can be written in the form
\begin{eqnarray}
\left|\lambda_0\right|&=&
\sqrt{\prod _{k=0}^{2L-2}
\left[1+  \sin (4 u)
\sin \left(\frac{\pi k}{2 L-1}\right)\right]} . \nonumber
\end{eqnarray}
The argument $\theta_0$ of the largest eigenvalue $\lambda_0$ is given by
\begin{eqnarray}
\theta_0=2u-\frac{\pi}{4}+\sum_{k=1}^{L-1}\arctan{\frac{\cos(4u)\cos{\left(\frac{\pi  (2 k-1)}{2 L-1}\right)}}{\sin(4u)+\sin\left(\frac{\pi  (2 k-1)}{2 L-1}\right)}} \label{alpha1} .
\end{eqnarray}
Let us change variable $k$ as $k \to L-k$.  Then the argument $\theta_0$ given by Eq. (\ref{alpha1}) can be transformed to the form
\begin{eqnarray}
\theta_0=2u-\frac{\pi}{4}-\sum_{k=1}^{L-1}\arctan{\frac{\cos(4u)\cos{\left(\frac{2\pi k}{2 L-1}\right)}}{\sin(4u)+ \sin\left(\frac{2\pi k}{2 L-1}\right)}} \label{alpha2} .
\end{eqnarray}
The sum over $k$ in Eq. (\ref{alpha2}) can be extended up to $2 L - 2$ as
\begin{eqnarray}
\theta_0&=&2u-\frac{\pi}{4}+\frac{1}{2}\sum_{k=1}^{2L-2}(-1)^{k+1}\arctan{\frac{\cos(4u)\cos{\left(\frac{\pi k}{2 L-1}\right)}}{\sin(4u)+\sin\left(\frac{\pi k}{2 L-1}\right)}} \label{alpha3}\\
&=&\frac{1}{2}\sum_{k=0}^{2L-2}(-1)^{k+1}\arctan{\frac{\cos(4u)\cos{\left(\frac{\pi k}{2 L-1}\right)}}{\sin(4u)+ \sin\left(\frac{\pi k}{2 L-1}\right)}} . \label{alpha4}
\end{eqnarray}
In the last step we have use the fact that $\frac{1}{2}\arctan{\cot{(4u)}}=\frac{\pi}{4}-2u$.

The derivation of the asymptotic expansion of $\log \lambda_0 = \log \left|\lambda_0\right| + i \theta_0$ can be divided to two parts. First,  with the help of the Euler-Maclaurin summation formula (see Eq. (\ref{general EM}) in appendix A) we can derive the asymptotic expansion of the logarithm of the absolute value of $\lambda_0$
\begin{eqnarray}
\label{Abs[L_0]}
\log \left|\lambda_0\right|&=&\frac{1}{2}\sum_{k=0}^{2L-2}\log \left[1+\sin (4 u)
\sin \left(\frac{\pi k}{2 L-1}\right)\right]\nonumber\\
&=& -f_{\rm{bulk}}(2L-1)-2\sum_{k=0}^{\infty}\frac{B_{2k+2}f^{(2k+1)}(0)}{(2k+2)!}\left(\frac{\pi}{2L-1}\right)^{2k+1}\label{lambda0Exp}\\
&=& - f_{\rm{bulk}}(2L-1)-\frac{\pi \sin(4u)}{12}\frac{1}{2L-1}-\frac{\pi^3 \sin(4u) \cos(8u)}{720 (2L-1)^3}+...
\end{eqnarray}
where $f(x)$ is given by
\begin{equation}
f(x)=\frac{1}{2}\log\left[1+\sin(4u) \sin x \right]
\end{equation}
and $f_{\rm{bulk}}$ is given by
\begin{equation}
f_{\rm{bulk}}=-\frac{1}{\pi}\int_{0}^{\pi} f(x) dx=-\frac{1}{2\pi}\int_{0}^{\pi}\log\left[1+\sin(4u) \sin x \right]dx.
\end{equation}
Now with the help of the Boole summation formula (see Eq. (\ref{Boole}) in appendix A) the asymptotic expansion of the argument $\theta_0$ can be written in the form
\begin{eqnarray}
\theta_0&=&\frac{1}{2}\sum_{k=0}^{2L-2}(-1)^{k+1}\arctan{\frac{\cos(4u)\cos{\left(\frac{\pi k}{(2 L-1)}\right)}}{\sin(4u)+\sin\left(\frac{\pi k}{(2 L-1)}\right)}} \nonumber\\
&=&-\frac{1}{2}\sum_{n=0}^{\infty}\frac{E_{2n+1}(0)}{(2n+1)!}
\left[g^{(2n+1)}(\pi)+g^{(2n+1)}(0)\right]\left(\frac{\pi}{2L-1}\right)^{2n+1}
\end{eqnarray}
where function $g(x)$ is given by
\begin{equation}
g(x)=\frac{1}{2}\arctan{\frac{\cos(4u)\cos x}{\sin(4u)+\sin x}}.
\end{equation}
Since the function $g(x)$ obeys the symmetry $g(x)=-g(\pi-x)$, the derivatives of the function $g(x)$ at the points zero and $\pi$ are related to each other as
\begin{eqnarray}
g^{(2k+1)}(0)&=&g^{(2k+1)}(\pi), \qquad \qquad k=0,1,2,...\label{godd}\\
g^{(2k)}(0)&=&-g^{(2k)}(\pi), \qquad \qquad k=0,1,2,....\label{geven}
\end{eqnarray}
Using Eq. (\ref{godd}) the asymptotic expansion of the argument of $\lambda$  can finally be written in the form
\begin{eqnarray}
\theta_0&=&-\sum_{n=0}^{\infty}\frac{E_{2n+1}(0)g^{(2n+1)}(0)}{(2n+1)!}
\left(\frac{\pi}{2L-1}\right)^{2n+1}\label{theta1}\\
&=&-\frac{\pi \cos{(4u)}}{4(2L-1)}+\frac{\pi^3 \sin(4u) \sin(8u)}{48 (2L-1)^3}+... .
\end{eqnarray}
Thus the leading finite-size corrections to the ground state energies ${\cal E}_0=\frac{\sqrt{2}}{2} {\bf E}_0 =-\frac{\sqrt{2}}{2}\log \lambda_0$ take the form
\begin{eqnarray}
\label{log l_0}
{\cal E}_0=-\frac{\sqrt{2}}{2}\log{\lambda_0}&=&f_{\rm{bulk}}\frac{\sqrt{2}}{2} (2L-1)+\frac{2\sqrt{2}\pi}{2L-1}
\left(\frac{1}{48}\sin (4 u)
+\frac{i}{16}\cos (4 u)\right)+O\left(\frac{1}{2L-1}\right).\nonumber\\ \label{E0Ising}
\end{eqnarray}
The leading finite-size corrections to the ground state energies ${\cal E}_0$ given by Eq. (\ref{energy3}) for the Ising model with $c=1/2$ and $g=4$  take the form
\begin{eqnarray}
{\cal E}_0&=&L_x f_{\rm{bulk}}+\frac{2\pi}{L_x}\left[\left(-\frac{1}{24}+\Delta+\bar\Delta\right)\sin(4 u) + i(\Delta-\bar\Delta)\cos(4u) \right]+O\left(\frac{1}{L_x}\right)
\label{E01I}\\
&=&L_x f_{\rm{bulk}}+\frac{2\pi i}{L_x}\left[\left(-\frac{1}{48}+\Delta\right)e^{-i 4 u}-\left(-\frac{1}{48}+\bar\Delta\right)e^{i 4 u} \right]+O\left(\frac{1}{L_x}\right)\label{E02I}.
\end{eqnarray}
Compare Eqs. (\ref{E0Ising}) and (\ref{E01I}) for $L_x=\sqrt{2}(L-1/2)$ we can see that $\Delta = 1/16$ and $\bar \Delta = 0$. Thus we have shown that duality twisted boundary conditions with $\left(\Delta, \bar \Delta\right) = \left(\frac{1}{16}, 0\right)$ correspond to the boundary conditions with seam $(r,s)=(1,2)$.

Let us now consider the other eigenvalues given by Eq. (\ref{l new}), which correspond to some combination of the $\{\mu_k\}=(\mu_1,\mu_2,...,\mu_{L-1})$.  Denote by $\lambda_{\{p\}}$ the eigenvalue given by Eq. (\ref{l new}) with $\mu_p=-1$ and the other $\mu_k$ have value $\mu_k=+1$. The first exited state corresponds to the second largest eigenvalues $\lambda_1=\lambda_{\{1\}}$, and the second exited state corresponds to the third largest eigenvalue $\lambda_2=\lambda_{\{L-1\}}$. It is easy to show that
\begin{eqnarray}
\lambda_{\{p\}}&=&\lambda_0\left[\frac{\exp[2i\left(u-\frac{\pi}{8}\right)]\sin\left(\frac{\pi(2p-1)}{4L-2}\right)
-\exp[-2i\left(u-\frac{\pi}{8}\right)]\cos\left(\frac{\pi(2p-1)}{4L-2}\right)}{\exp[2i\left(u-\frac{\pi}{8}\right)]
\sin\left(\frac{\pi(2p-1)}{4L-2}\right)
+\exp[-2i\left(u-\frac{\pi}{8}\right)]\cos\left(\frac{\pi(2p-1)}{4L-2}\right)}\right]^2
\label{Lp}\\
&=&\lambda_0 A_p e^{i \phi_p} \label{Lp1}
\end{eqnarray}
where $A_p$ and $\phi_p$ are given by
\begin{eqnarray}
\log A_p&=&\log \frac{1-\sin(4u)\sin\left(\frac{\pi(2p-1)}{2L-1}\right)}
{1+\sin(4u)\sin\left(\frac{\pi(2p-1)}{2L-1}\right)}
=-4
\sum_{n=0}^{\infty}
\frac{(2p-1)^{2n+1}f^{(2n+1)}(0)}{(2n+1)!}
\left(\frac{\pi}{2L-1}\right)^{2n+1}
\label{Amod}\\
\phi_p&=&2\arctan\left[\cos(4u)\tan\left(\frac{\pi(2p-1)}{2L-1}\right)\right]
=-4
\sum_{n=0}^{\infty}
\frac{(2p-1)^{2n+1}g^{(2n+1)}(0)}{(2n+1)!}
\left(\frac{\pi}{2L-1}\right)^{2n+1}
\label{phi}
\end{eqnarray}
For values of $p$ close to $L$ its convenient consider instead of Eq. (\ref{Lp1})  the  expressions
\begin{eqnarray}
\lambda_{\{L-p\}}&=&\lambda_0 A_{L-p} e^{i \phi_{L-p}} . \label{Lp2}
\end{eqnarray}
From Eqs. (\ref{Amod}) and (\ref{phi})  one can easily obtain the expressions for the $A_{L-p}$ and $\phi_{L-p}$:
\begin{eqnarray}
\log A_{L-p}&=&\log \frac{1-\sin(4u)\sin\left(\frac{2 p \pi}{2L-1}\right)}{1+\sin(4u)\sin\left(\frac{2 p \pi}{2L-1}\right)}
=-4
\sum_{n=0}^{\infty}
\frac{(2p)^{2n+1}f^{(2n+1)}(0)}{(2n+1)!}
\left(\frac{\pi}{2L-1}\right)^{2n+1}
 \label{Amod1}\\
\phi_{L-p}&=&-2\arctan\left[\cos(4u)\tan\left(\frac{2 p \pi}{2L-1}\right)\right]
=4
\sum_{n=0}^{\infty}
\frac{(2p)^{2n+1}g^{(2n+1)}(0)}{(2n+1)!}
\left(\frac{\pi}{2L-1}\right)^{2n+1}.
 \label{phi1}
\end{eqnarray}

The asymptotic expansions of  $\log\lambda_{\{p\}}$ and $\log\lambda_{\{L-p\}}$ can be written by using Eqs. (\ref{theta}), (\ref{lambda0Exp}), (\ref{Amod}), (\ref{phi}), (\ref{Amod1}) and (\ref{phi1}) as
\begin{eqnarray}
-\log\lambda_{\{p\}}&=&f_{\rm{bulk}}(2L-1)+
2\sum_{n=0}^{\infty}
\left(
\frac{B_{2n+2}}{2n+2}+2(2p-1)^{2n+1}
\right)
\frac{f^{(2n+1)}(0)}{(2n+1)!}
\left(
\frac{\pi}{2L-1}
\right)^{2n+1}\nonumber\\
&&+i\sum_{n=0}^{\infty}
\left(E_{2n+1}(0)+4(2p-1)^{2n+1}\right)
\frac{g^{(2n+1)}(0)}{(2n+1)!}
\left(\frac{\pi}{2L-1}\right)^{2n+1}\label{EStae1}\\
-\log\lambda_{\{L-p\}}&=&f_{\rm{bulk}}(2L-1)
+2\sum_{n=0}^{\infty}
\left(
\frac{B_{2n+2}}{2n+2}+2(2p)^{2n+1}
\right)
\frac{f^{(2n+1)}(0)}{(2n+1)!}
\left(
\frac{\pi}{2L-1}
\right)^{2n+1}\nonumber\\
&&+i\sum_{n=0}^{\infty}
\left(E_{2n+1}(0)-4(2p)^{2n+1}\right)
\frac{g^{(2n+1)}(0)}{(2n+1)!}
\left(\frac{\pi}{2L-1}\right)^{2n+1} \label{EStae2}.
\end{eqnarray}
From Eqs. (\ref{EStae1}) and (\ref{EStae2}) one can obtain the leading finite-size corrections to the exited state energies ${\cal E}_{p}=\frac{\sqrt{2}}{2} {\bf E}_{p} =-\frac{\sqrt{2}}{2}\log \lambda_{\{p\}}$  and ${\cal E}_{L-p}=\frac{\sqrt{2}}{2} {\bf E}_{L-p} =-\frac{\sqrt{2}}{2}\log \lambda_{\{L-p\}}$ in the form
\begin{eqnarray}
&&{\cal E}_{p}=-\frac{\sqrt{2}}{2}\log \lambda_{\{p\}}=L_x f_{\rm{bulk}}\nonumber\\
&&+\frac{2\pi}{L_x}
\left[\left(-\frac{23}{48}+p\right)\sin{(4u)}
+i \left(\frac{9}{16}-p\right)\cos{(4u)}
\right]+O\left[\frac{1}{L_x}\right]^2\nonumber\\
&&= L_x f_{\rm{bulk}}+\frac{2\pi i}{L_x}
\left[\frac{1}{24}e^{-4iu}
+\left(\frac{25}{48}-p\right)e^{4iu}
\right]+O\left[\frac{1}{L_x}\right]^2\label{Epf}\\
&&{\cal E}_{L-p}=-\frac{\sqrt{2}}{2}\log \lambda_{\{L-p\}}=L_x f_{\rm{bulk}}\nonumber\\
&&+\frac{2\pi}{L_x}
\left[\left(\frac{1}{48}+p\right)\sin(4u)
+i \left(\frac{1}{16}+p\right)\cos(4u)
\right]+O\left[\frac{1}{L_x}\right]^2 \nonumber\\
&&=L_x f_{\rm{bulk}}+\frac{2\pi i}{L_x}
\left[\left(\frac{1}{24}+p\right)e^{-4iu}
+\frac{1}{48}e^{4iu}
\right]+O\left[\frac{1}{L_x}\right]^2 \label{Epf1}
\end{eqnarray}
where $L_x$ is given by Eq. (\ref{Lx}).

The leading finite-size corrections to the exited state energies ${\cal E}_n$  given by Eqs. (\ref{energy}) and (\ref{energy1}) for the Ising model with $c=1/2$ and $g=4$  take the form
\begin{eqnarray}
{\cal E}_n&=&L_x f_{\rm{bulk}}+\frac{2\pi}{L_x}\left[\left(-\frac{1}{24}+\Delta+\bar\Delta+k_n+\bar k_n\right)\sin(4 u) + i(\Delta-\bar\Delta+k_n-\bar k_n)\cos(4u) \right]\nonumber\\
&+&O\left(\frac{1}{L_x}\right)
\nonumber\\
&=&L_x f_{\rm{bulk}}+\frac{2\pi i}{L_x}\left[\left(-\frac{1}{48}+\Delta+k_n\right)e^{-i 4 u}-\left(-\frac{1}{48}+\bar\Delta+\bar k_n\right)e^{i 4 u} \right]+O\left(\frac{1}{L_x}\right).\label{E02I2}
\end{eqnarray}
Now compare Eqs. (\ref{Epf}),  (\ref{Epf1})  with Eq. (\ref{E02I2})  we can see that for the exited state ${\cal E}_{p}$ we have
\begin{equation}
|\Delta = 1/16, k_p=0; \bar \Delta = 1/2, \bar k_p = p-1\rangle
\label{statep}
\end{equation}
and for the exited state ${\cal E}_{L-p}$ we have
\begin{equation}
|\Delta = 1/16, k_{L-p}=p; \bar \Delta = 0, \bar k_{L-p} = 0\rangle.
\label{stateLp}
\end{equation}

\section{Twisted partition function $Z_{(1,2)}(p)$}
\label{partfunc}
Now we have all the necessary information to start the calculations of partition function for the Ising model with duality-twisted boundary conditions $Z_{L_xL_y}$. For large $L_x$ and $L_y$ (always keeping the ratio $L_y/L_x$ constant) we have
\begin{eqnarray}
\label{Z}
Z_{L_x L_y}=\sum_l\lambda_l^M\approx e^{-L_xL_yf_{\rm{bulk}}}Z_{(1,2)}(q),
\end{eqnarray}
where $f_{\rm{bulk}}$ is the bulk free energy, $q$ is the modular parameter and $Z_{(1,2)}(q)$ is the universal conformal partition function.

Let us now find the general form of the eigenvalues with significant input
in the partition function. From Eq. (\ref{abs}) we can see that these are eigenvalues
for which almost all $\mu_k=1$ and some $\mu_k$ are allowed to take the value $-1$
only if $k\ll L$ or $L-k\ll L$. Any ``significant'' eigenvalue will be specified by two sets of indexes $K=\{k_1,k_2,\cdots,k_m\}$ and
 $\bar K=\{\bar k_1,\bar k_2,\cdots,\bar k_{\bar m}\}$, where $\bar k_i \equiv L-k_i$ with
 $k_1< k_2<k_3<\cdots<k_m\ll L $ and $\bar{ k_1}<\bar{k_2}<\bar{k_3}<\cdots
<\bar{k}_{\bar{ m}}\ll L$ so that $\mu_{\bar k_i}=-1$,$\mu_{k_i}=-1$ and the other
$\mu$'s are $+1$. From above it is easy  to get
\begin{eqnarray}
\label{log l}
&&-\frac{\sqrt 2}{2}\log\lambda_{K\bar K}=L_x f_{\rm{bulk}}\nonumber\\
&&+\frac{2\pi i}{L_x}
\left[ \left( -\frac {c}{24} + \frac {1}{16} + \sum_i k_i\right)e^{-4iu}-
\left(-\frac{c}{24}+\sum_i \left(\bar{k}_i -\frac {1}{2}\right)\right) e^{4iu}\right]+O\left[\frac{1}{N}\right]^2.\label{Holo}
\end{eqnarray}
\red{From antiholomorphic part of the Eq. (\ref{Holo}) one can see that exited state in antiholomorphic region is correspond to the conformal state $\bar \psi_{-\bar \nu_1}\bar \psi_{-\bar \nu_2}...\bar\psi_{-\bar \nu_n}|0>$, where $\bar \nu_i = \bar k_i-1/2$ are half integer numbers. In particular  the exited state ${\cal E}_{p}$ ($|p>$) given by Eq. (\ref{statep}) corresponds to the conformal state $\bar\psi_{-p+1/2}|0>$. It is easy to show that 
\begin{equation}
[L_{-1}\bar \psi_{m}]=\left(\frac{1}{2}-m\right)\,\bar\psi_{m-1} 
\end{equation}
from which, using the fact that $L_{-1}|0>=0$ one can derive that 
\begin{equation}
\bar\psi_{-p+1/2}|0>=\frac{1}{(p-1)!}[L_{-1}^{p-1}\bar \psi_{-1/2}]|0>=\frac{1}{(p-1)!}L_{-1}^{p-1}\bar \psi_{-1/2}|0>=\frac{1}{(p-1)!}L_{-1}^{p-1}|1/2>
\end{equation}
Thus the state $|p>$ is the descendent conformal state $L_{-1}^{p-1}|\bar \Delta>$ generated by the primary field $\bar \psi(\bar z)$ with conformal dimension $\bar \Delta=1/2$. Now from holomorphic part of the Eq. (\ref{Holo}) one can see that exited state in holomorphic region is correspond to the conformal state $\psi_{-k_1}\psi_{-k_2}...\psi_{-k_n}|\sigma>$, where $|\sigma>=\sigma(0)|0>$ is the asymptotic state created by a primary field $\sigma(0)$ of the conformal dimension $\Delta = 1/16$. The exited state ${\cal E}_{p}$ ($|L-p>$) given by Eq. (\ref{stateLp}) correspond to the conformal state $\psi_{-p}|\sigma>=\psi_{-p}\sigma(0)|0>$, which can be considered as descendent of the conformal state generated by product of two conformal fields $\psi(z)\sigma(0)$.}

Now from Eqs. (\ref{Z}) and (\ref{log l}) we can obtain that
\begin{eqnarray}
\label{UnPart}
Z_{(1,2)}(q)=\sum_{K\bar K}q^{-\frac{c}{24}+\frac{1}{16}+\sum_i k_{i}}
\bar{q}^{-\frac{c}{24}+\sum_i\left(\bar{k}_{i}-\frac{1}{2}\right)}
\end{eqnarray}
where $q$ is a modular parameter given by
\begin{eqnarray}
\label{mod par}
q=e^{-\frac{2 \pi i L_y}{L_x} e^{-4 iu}}.
\end{eqnarray}
For further calculations it is convenient to introduce the occupation numbers
$\varepsilon_k=\frac{1-\mu_k}{2}$ and $\bar{\varepsilon}_{\bar{k}}=\frac{1-\bar\mu_{\bar k}}{2}$.
In terms of this quantities the universal conformal partition function Eq. (\ref{UnPart}) can be
rewritten as
\begin{eqnarray}
\label{UnPart2}
&&Z_{(1,2)}(q)=\sum_{\{\varepsilon\}\{\bar\varepsilon\}}q^{-\frac{c}{24}+\frac{1}{16}
+\sum_k^\infty k \varepsilon_{k}}
\bar{q}^{-\frac{c}{24}+\sum_{\bar{k}}^{\infty}\left(\bar{k}-\frac{1}{2}\right)\bar
{\varepsilon}_{\bar{k}}}\nonumber\\
&&=q^{-\frac{c}{24}+\frac{1}{16}}\bar{q}^{-\frac{c}{24}}\sum_{\{\varepsilon\}\{\bar\varepsilon\}}\prod_{k=1}^{\infty}
q^{k\varepsilon_k}\prod_{\bar{k}=1}^{\infty}\bar{q}^{\left(\bar{k}-\frac{1}{2}\right)
\bar{\varepsilon}_{\bar{k}}}\nonumber\\
&&=q^{-\frac{c}{24}+\frac{1}{16}}\bar q^{-\frac{c}{24}}\sum_{\varepsilon_1\in\{0,1\}}\sum_{\varepsilon_2\in\{0,1\}}\cdots
\sum_{\bar\varepsilon_1\in\{0,1\}}\sum_{\bar\varepsilon_2\in\{0,1\}}\cdots
\prod_{k=1}^\infty q^{k\varepsilon_k}
\prod_{\bar k=1}^\infty \bar q^{\left(\bar k-\frac{1}{2}\right)\bar\varepsilon_{\bar k}}
\nonumber\\
&&=q^{-\frac{c}{24}+\frac{1}{16}}\bar q^{-\frac{c}{24}}\prod_{k=1}^\infty \left(1+q^k\right)
\prod_{\bar k=1}^\infty\left(1+\bar q^{\bar k -\frac{1}{2}} \right ).
\end{eqnarray}
Taking into account that
\begin{eqnarray}
\label{VC}
&&\chi_0=\frac{1}{2}q^{-\frac{1}{48}}\left
(\prod _{k=1}^{\infty } \left(1+q^{k-\frac{1}{2}}\right)
+\prod _{k=1}^{\infty } \left(1-q^{k-\frac{1}{2}}\right)\right)\nonumber\\
&&\chi _{\frac{1}{2}}=\frac{1}{2}q^{-\frac{1}{48}}\left
(\prod _{k=1}^{\infty } \left(1+q^{k-\frac{1}{2}}\right)
-\prod _{k=1}^{\infty } \left(1-q^{k-\frac{1}{2}}\right)\right)\nonumber\\
&&\chi _{\frac{1}{16}}=q^{\frac{1}{24}}
\prod _{k=1}^{\infty }\left(1+q^{k}\right),
\end{eqnarray}
and adding the conjugate part of the eigenvalue set we get
\begin{eqnarray}
\label{Z with VC}
Z_{(1,2)}(q)=\left[\chi _0(q)+\chi _{\frac{1}{2}}(q)\right]^*\chi _{\frac{1}{16}}(q)+
\left[\chi _0(q)+\chi _{\frac{1}{2}}(q)\right]\chi _{\frac{1}{16}}(q)^*.
\end{eqnarray}
Thus we have obtained analytically the universal conformal partition function for the Ising model with duality-twisted boundary conditions $Z_{(1,2)}(q)$ which confirms the numerical result of \cite{Pearce2001}.

\section{Universal amplitude ratios}
\label{univratio}
In this section we present the set of universal amplitude ratios for finite-size corrections of the two-dimensional Ising model on square lattices with duality-twisted boundary conditions given by Eq. (\ref{eq7.4}). Let us denote the free energy per spin $f$, the inverse correlation lengths $\xi^{-1}_{p}$ and $\xi^{-1}_{L-p}$  of our critical Ising model as
\begin{eqnarray}
L_x f&=&-\frac{\sqrt{2}}{2}\log\left(\left|\lambda_0\right|\right)\label{F1}\\
\xi^{-1}_{p}&=&\frac{\sqrt{2}}{2}\log\left(\left|\frac{\lambda_0}{\lambda_{\{p\}}}\right|\right)=
-\frac{\sqrt{2}}{2}\log{A_{p}}\label{B}\\
\xi^{-1}_{L-p}&=&\frac{\sqrt{2}}{2}\log\left(\left|\frac{\lambda_0}{\lambda_{\{L-p\}}}\right|\right)=
-\frac{\sqrt{2}}{2}\log{A_{L-p}}. \label{C}
\end{eqnarray}
From Eqs. (\ref{Abs[L_0]}), (\ref{Amod}) and (\ref{Amod1}) it is clear that asymptotic expansions for the free energy per spin $f$ and the inverse correlation lengths $\xi^{-1}_{p}$ and $\xi^{-1}_{L-p}$  can be written in the forms
\begin{eqnarray}
L_x (f-f_{\rm{bulk}})&=&\sum_{k=1}^{\infty}\frac{a_{k}(p)}{L_x^{2k-1}}\label{F2}\\
\xi^{-1}_{p}&=&\sum_{k=1}^{\infty}\frac{b_{k}(p)}{L_x^{2k-1}}\label{B1}\\
\xi^{-1}_{L-p}&=&\sum_{k=1}^{\infty}\frac{c_{k}(p)}{L_x^{2k-1}}, \label{C1}
\end{eqnarray}
where coefficients $a_{k}(p)$, $b_{k}(p)$ and $c_{k}(p)$ are given by
\begin{eqnarray}
a_{k}(p)&=&\frac{\pi^{2k-1}B_{2k}}{2^{k-1}(2k)!}f^{(2k-1)}(0) \label{a}\\
b_{k}(p)&=&\frac{\pi^{2k-1}(2p-1)^{2k-1}}{2^{k-2}(2k-1)!}f^{(2k-1)}(0)  \label{b}\\
c_{k}(p)&=&\frac{2^{k+1}\pi^{2k-1}p^{2k-1}}{(2k-1)!}f^{(2k-1)}(0).  \label{c}
\end{eqnarray}
The coefficients $a_1(p)$, $b_1(p)$, $c_1(p)$
\begin{eqnarray}
a_1(p)&=&\frac{\pi}{24}\sin{y}\label{a10}\\
b_1(p)&=&(2p-1) \pi \sin{y}\label{b10}\\
c_1(p)&=&2 p \pi \sin{y} \label{c10}
\end{eqnarray}
are universal and related  to the conformal anomaly number (c), the conformal dimensions of the ground state ($\Delta, \bar \Delta$), and the scaling dimensions
of the $p$-th scaling fields ($x_p, z_p$) of the theory in the following way
\begin{eqnarray}
a_1(p)&=&2\pi \zeta \left(-\frac{c}{12}+\Delta+\bar \Delta\right)\label{a1}\\
b_1(p)&=&2\pi \zeta x_p \label{b1}\\
c_1(p)&=&2\pi \zeta z_p, \label{c1}
\end{eqnarray}
with $c=1/2$, $\Delta=1/16$, $\bar \Delta=0$, $x_p=p-1/2$, $z_p=p$ and anisotropy parameter $\zeta=\sin y$.

The coefficients $a_{k}(p)$, $b_{k}(p)$ and $c_{k}(p)$ for $k \ge 2$ are non-universal, since they depends on non-universal parameter $f^{(2k-1)}(0)$, but ratio of this coefficients are universal and given by
\begin{eqnarray}
r_p(k)=\frac{b_{k}(p)}{a_{k}(p)}&=&\frac{4 k(2p-1)^{2k-1}}{B_{2k}}\label{r}\\
r_{L-p}(k)=\frac{c_{k}(p)}{a_{k}(p)}&=&\frac{4 k(2p)^{2k-1}}{B_{2k}}. \label{Lr}
\end{eqnarray}
Below we will show that universality of the ratios $r_p(k), r_{L-p}(k)$ for $k=1$ and $2$  follow from the conformal field theory.

The case for $k=1$ is trivial, since $r_p(1)$ and $r_{L-p}(1)$ are the ratios of universal coefficients $a_{1}(p)$, $b_{1}(p)$ and  $c_{1}(p)$ and equal to
\begin{eqnarray}
r_p(1) &=& 24 (2p-1)\label{r1}\\
r_{L-p}(1)&=& 48 p. \label{rL1}
\end{eqnarray}
The case $k=2$ is non-trivial. The ratios $r_p(2)$ and $r_{L-p}(2)$ are given by
\begin{eqnarray}
r_p(2) &=& -240 (2p-1)^3\label{r2}\\
r_{L-p}(2)&=& -1920 p^3 . \label{rL2}
\end{eqnarray}

\section{Perturbative conformal field theory}
\label{pcft}
The finite-size corrections to Eq. (\ref{energy}) can in principle be computed
in perturbative conformal field theory. In general, any critical lattice Hamiltonian
 contains correction terms to the fixed-point Hamiltonian $H_c=\frac{\pi}{L_x}\left(L_0+\bar L_0-\frac{c}{12}\right)$
\begin{equation}
H =\zeta H_c + \sum_k g_k \int_{-L_x/2}^{L_x/2}\phi_k(v) d v, \label{Hc}
\end{equation}
where $g_k$ is a non-universal constant and $\phi_k(v)$ is a
perturbating conformal field with scaling dimension $x_k$. Among
these fields are those associated with the conformal block of the
identity operator, the leading operator of which has the scaling
dimension $x_l=4$. To  first order in the perturbation, the
energy gaps $({\cal E}_n-{\cal E}_0)$ and the ground-state energy (${\cal E}_0$) can be
written as
\begin{eqnarray}
{\cal E}_n-{\cal E}_0&=&\frac{2 \pi}{L_x}\zeta x_n+ 2 \pi \sum_k
g_k(C_{nkn}-C_{0k0})\left(\frac{2 \pi}{L_x}\right)^{x_k-1},
\nonumber\\
{\cal E}_0 &=& {\cal E}_{0,c}+2 \pi \sum_k g_k C_{0k0} \left(\frac{2
\pi}{L_x}\right)^{x_k-1}, \nonumber
\end{eqnarray}
where $C_{nkn}$ are universal structure constants. Note, that the
ground state energy ${\cal E}_0$ and the energy gaps (${\cal E}_n-{\cal E}_0$) are, respectively, the quantum analogues of the free energy $f$
and inverse spin-spin correlation lengths $\xi_n^{-1}$; that is, $L_x
f \Leftrightarrow  {\cal E}_0, \quad \xi_n^{-1} \Leftrightarrow
{\cal E}_n-{\cal E}_0$.

In the case of the cylinder
geometry the spectra of the Hamiltonian (\ref{Hc})  are built by
the irreducible representation $\Delta, \bar \Delta$ of two
commuting Virasoro algebras $L_n$ and ${\bar L}_n$.
The leading finite-size
corrections ($1/L_x^3$) can be described by the Hamiltonian given by
Eq. (\ref{Hc}) with a single perturbating conformal field
$\phi_1(v)=L_{-2}^2(v)+\bar L_{-2}^2(v)$ with scaling dimension $x_1=4$, which
belongs to the tower of the identity  \cite{henkel}. Thus the ratio $r_n(2)$ are indeed universal and given by
\begin{equation}
r_n(2)=\frac{C_{n1n}-C_{010}}{C_{010}}.
\nonumber
\end{equation}

The universal structure constants $C_{n1n}$  can be
obtained from the matrix elements $\langle n|\phi_1(0)|n \rangle
=\left({2 \pi}/{L_x}\right)^{x_1}C_{n1n}$ \cite{cardy86}, which for descendent states ($L_{-1}^n|\Delta\rangle$) generated by primary field $\psi(z,\bar z)$ with conformal dimension $\Delta$ have already been computed by Reinicke \cite{reinicke87}:
\begin{eqnarray}
C_{n1n}&=&\left(\frac{c}{24}\right)^2+\frac{11 c}{1440} + (\Delta+r)\left(\Delta-\frac{2+c}{12}+\frac{r(2 \Delta +
r)(5 \Delta+1)}{(\Delta+1)(2\Delta+1)}\right)
\nonumber\\
&+&\frac{r}{30}
 \left[r^2(5c-8)-(5c+28)\right]\delta_{\Delta,0}+ (\Delta \rightarrow \bar \Delta, r
\rightarrow \bar r).\label{cnlnd0}
\end{eqnarray}
Let us consider the case $c=1/2$. For the two-dimensional Ising model with duality-twisted boundary conditions the ground state $|0\rangle$, the excited states $|p\rangle$ and $|L-p\rangle$ are given by
\begin{eqnarray}
|0\rangle&=&|\Delta_0=\frac{1}{16},r=0;\bar \Delta_0=0,\bar
r=0\rangle \label{Es0}\\
|p\rangle&=&|\Delta_p=\frac{1}{16},r=0;\bar \Delta_p=\frac{1}{2},\bar r=p-1\rangle \label{Esp}\\
|L-p\rangle&=&|\Delta_{L-p}=\frac{1}{16},r=p;\bar \Delta_{L-p}=0,\bar r=0\rangle . \label{EsLp}
\end{eqnarray}
Note that the excited states $|p\rangle$ can be identified as descendent states of the primary field $\psi(z)$ with conformal dimension $\bar \Delta = 1/2$, while the excited states $|L-p\rangle$ can be identified as descendent states generated by the OPE of the primary fields $\psi(z)\sigma(0)$ where $\sigma(0)$ is the spin operator with conformal dimension $\Delta = 1/16$.
Thus for descendent states of the primary field the universal structure constants $C_{n1n}$ can be obtained from Eq. (\ref{cnlnd0}):
\begin{eqnarray}
C_{010}&=&-\frac{7}{11520}, \label{C010}\\
C_{p1p}&=& -\frac{7}{11520}+\frac{7(2p-1)^3}{48}. \label{Cp1p}
\end{eqnarray}
Now the ratios $r_p$ are given by
\begin{eqnarray}
r_p&=&\frac{C_{p1p}-C_{010}}{C_{010}}= -240(2p-1)^3\label{rpconf}
\end{eqnarray}
which is exactly coincide with Eq. (\ref{r2}) for all $p$. Thus we have obtained from CFT that ratios $r_p$ indeed universal and given by Eqs. (\ref{r2}).

Naive application of Eq. (\ref{cnlnd0}) for the universal structure constants $C_{(L-p)1(L-p)}$ would lead to
\begin{equation}
C_{(L-p)1(L-p)}=-\frac{7}{11520}-\frac{7p}{51}+\frac{7p^2}{34}+\frac{56p^3}{51}\label{CLp1Lp},
\end{equation}
and for $r_{L-p}$ we obtain
\begin{eqnarray}
r_{L-p}&=&\frac{C_{(L-p)1(L-p)}-C_{010}}{C_{010}}=-\frac{1920}{17}p(16p^2+3p-2)\label{rLpconf}
\end{eqnarray}
which coincides with Eq. (\ref{rL2}) only for $p = 1$ and $2$. For $p = 1$ we have only one descendant field $L_{-1}$ for which we can apply the Reinicke formula for descendent states ($L_{-1}^n|\Delta\rangle$) given by Eq. (\ref{cnlnd0}). For $p=2$ we have two descendent fields $L_{-1}^2$ and $L_{-2}$, but since at level 2 we have null vector $L_{-2}+\eta L_{-1}^2$ there is actually only one independent descendent field $L_{-1}^2$ and again we can apply the Reinicke formula for ($L_{-1}^n|\Delta\rangle$) states. The situation with $p \ge 3$ is drastically different since excited states $|L-p\rangle$ for $p \ge 3$ can be identified as descendent states generated by the OPE of primary fields $\psi(z)\sigma(0)$ for which Reinicke formula is invalid and as result the Eq. (\ref{rLpconf}) is different from Eq. (\ref{rL2}). For such states the calculations of the universal structure constants $C_{n1n}$ is not straightforward since it involves knowledge of the four-point correlation function. The conformal invariance in general does not fix the precise form of the four-point correlation function but for some particular cases it is possible to write down explicit form of the four-point correlation function. One of such cases is Ising model and in the next section we have calculated for the Ising model the universal structure constants $C_{n1n}$ for descendent states generated by the OPE of the primary fields $\psi(z)\sigma(0)$ and find that the results are in complete agreement with Eq. (\ref{rL2}) for all values of $p$. Thus we have obtained from CFT that ratios $r_p$ and $r_{L-p}$ indeed universal and given by Eqs. (\ref{r2}) and (\ref{rL2}).

\section{Universal structure constants}
\label{Cnnn}
\subsection{Universal structure constants for descendent states generated by primary field $\psi(z,\bar z)$.}
Let us first reobtain the universal structure constants for descendent states ($L_{-1}^n|\Delta\rangle$) generated by primary field $\psi(z,\bar z)$ with conformal dimension $\Delta$ \cite{reinicke87}.
The universal structure constants $C_{n1n}$ can be obtained from the matrix elements
\begin{equation}
\langle n|\varphi_1(0,0)|n\rangle=\left(\frac{2\pi}{L_x}\right)^{x_1}C_{n1n},\label{phiknn}
\end{equation}
where $|n\rangle = |\Delta+r,\bar \Delta +\bar r\rangle$ is the exited state, $\varphi_1$ is the perturbating conformal field, which in our case is given by

\begin{equation}
\varphi_1(v,\bar v)=L_{-2}^2(v)+\bar L_{-2}^2(\bar v)\label{phik}
\end{equation}
with scaling dimension $x_1=4$.  The matrix element $\langle n|\varphi_1(0,0)|n\rangle$ can be obtained from the spectral decomposition of the three and two point correlation functions \cite{reinicke87}
\begin{equation}
\langle n|\varphi_1(0,0)|n\rangle=\left(\frac{2\pi}{L_x}\right)^{x_1}C_{n1n}=\frac{a_{r,r;\bar r,\bar r}}{b_{r,\bar r}}
\label{matrixelem}
\end{equation}
where $a_{r,r;\bar r,\bar r}$ and $b_{r,\bar r}$ are the coefficients in the spectral decomposition of the three and two point correlation functions under conformal mapping of the infinite plane (with coordinate z) to infinite cylinder of circumference $L_x$ (with coordinate $\omega$) $z=e^{\frac{2\pi}{L_x}\omega}$
\begin{eqnarray}
\langle 0|\psi(\omega_1,\bar \omega_1)\varphi_1(\omega_2, \bar \omega_2)\psi(\omega_3, \bar \omega_3)|0\rangle &=& \sum_{r_1,r_2;\bar r_1, \bar r_2} a_{r_1,r_2;\bar r_1, \bar r_2}\xi_1^{\Delta+r_1} \xi_2^{\Delta+r_2}\bar \xi_1^{\bar \Delta+\bar r_1}\bar \xi_2^{\bar \Delta+\bar r_2} \label{threepoint} \\
\langle 0|\psi(\omega_1,\bar \omega_1)\psi(\omega_3,\bar \omega_3)|0\rangle &=& \sum_{r,\bar r} b_{r,\bar r}\left(\xi_1 \xi_2\right)^{\Delta+r}\left(\bar \xi_1 \bar \xi_2\right)^{\bar \Delta+\bar r} \label{twopoint}
\end{eqnarray}
where $\Delta$ is conformal dimension of the primary field $\psi(\omega, \bar \omega)$. Also
\begin{equation}
\xi_i=\frac{z_i}{z_{i+1}}=\exp\left\{\frac{2\pi}{L_x}\left(\omega_i-\omega_{i+1}\right)\right\} \qquad \mbox{and} \qquad \xi_1 \xi_2=\frac{z_1}{z_3}
\label{xi}
\end{equation}
In what follow we will use the notation $\langle ... \rangle$ instead of $\langle 0|...|0\rangle$.
Since $\psi(\omega,\bar \omega)$ is a primary field which is transformed under the plane-cylinder mapping as
\begin{equation}
\psi(\omega,\bar \omega)=\left(\frac{2\pi}{L_x}\right)^{\Delta}\left(\frac{2\pi}{L_x}\right)^{\bar \Delta}z^{\Delta}{\bar z}^{\bar \Delta} \psi(z,\bar z),
\label{phiplcyl}
\end{equation}
the two point correlation function on the cylinder $\langle \psi(\omega_1,\bar \omega_1)\psi(\omega_3,\bar \omega_3)\rangle$ can be written as
\begin{equation}
\langle \psi(\omega_1,\bar \omega_1)\psi(\omega_3,\bar \omega_3)\rangle=\left(\frac{2\pi}{L_x}\right)^{2\Delta}\left(\frac{2\pi}{L_x}\right)^{2\bar \Delta}\left(z_1 z_3\right)^{\Delta}\left(\bar z_1 \bar z_3\right)^{\bar \Delta} \langle \psi(z_1,\bar z_1) \psi(z_3,\bar z_3)\rangle .
\label{twopointcyl}
\end{equation}
The two point correlation function on the infinite plane $\langle \psi(z_1,\bar z_1) \psi(z_3,\bar z_3)\rangle$ is known (see for example \cite{francesco} p. 180) and can be written as
\begin{equation}
\langle \psi(z_1,\bar z_1) \psi(z_3,\bar z_3)\rangle=\frac{1}{\left(z_1-z_3\right)^{2\Delta}}\frac{1}{\left(\bar z_1-\bar z_3\right)^{2\bar \Delta}} .
\label{twopointplane}
\end{equation}
After little algebra we can obtain that the two point correlation function on the cylinder $\langle \psi(\omega_1,\bar \omega_1)\psi(\omega_3,\bar \omega_3)\rangle$ can be written as
\begin{equation}
\langle \psi(\omega_1,\bar \omega_1)\psi(\omega_3,\bar \omega_3)\rangle=\left(\frac{2\pi}{L_x}\right)^{2\Delta}\left(\frac{2\pi}{L_x}\right)^{2\bar \Delta}\sum_{ r=0}^{\infty}\sum_{ \bar r=0}^{\infty}c_r c_{\bar r} \left(\xi_1\xi_2\right)^{\Delta+r}\left(\bar \xi_1 \bar \xi_2\right)^{\bar \Delta+\bar r} , \label{twopoint1}
\end{equation}
where coefficient $b_{r,\bar r}$ is given by
\begin{equation}
b_{r,\bar r}=\left(\frac{2\pi}{L_x}\right)^{2\Delta}\left(\frac{2\pi}{L_x}\right)^{2\bar \Delta}c_r c_{\bar r}
\label{brr}
\end{equation}
and
\begin{equation}
c_r=\frac{(2\Delta+r-1)!}{r!(2\Delta-1)!} .
\label{cr}
\end{equation}
Now let us calculate the three point correlation function on the cylinder
$\langle \psi(\omega_1,\bar \omega_1)\varphi_1(\omega_2, \bar \omega_2)\psi(\omega_3, \bar \omega_3)\rangle$
with $\varphi_1(\omega_2, \bar \omega_2)=L_{-2}^2(\omega_2)+\bar L_{-2}^2(\bar \omega_2)$. Using general transformation formula for conformal fields \cite{francesco,Gaberdiel1994} the conformal operator $L_{-2}^2(\omega)$ transforms under plane-cylinder mapping as
\begin{eqnarray}
L_{-2}^2(\omega)&=&\oint \oint \frac{T(\omega_1)T(\omega_2)}{(\omega_1-\omega)(\omega_2-\omega)}\frac{d \omega_1}{2\pi i}\frac{d \omega_2}{2\pi i}\nonumber\\
&=&\left(\frac{2\pi}{L_x}\right)^4\oint\oint\frac{\left[z_1^2 T(z_1)-\frac{c}{24}\right]\left[z_2^2 T(z_2)-\frac{c}{24}\right]}{z_1 z_2(\log{z_1}-\log{z})(\log{z_2}-\log{z})}\frac{d z_1}{2\pi i}\frac{d z_2}{2\pi i}\nonumber\\
&=& \left(\frac{2\pi}{L_x}\right)^{4}\left[z^4 L_{-2}^2(z)-\frac{c-10}{12}z^2 L_{-2}(z)+\frac{3 z^3}{2}L_{-3}(z)+\frac{\alpha c}{96}\right]
\label{L2transform}
\end{eqnarray}
where $\alpha=(22+5 c)/30$ and $c$ is the central charge. Here we have used the OPE of the energy-momentum tensor $T(z)$
\begin{equation}
T(z_1)T(z_2) \sim \frac{c/2}{(z_1-z_2)^4}+\frac{2T(z_2)}{(z_1-z_2)^2}+\frac{T'(z_2)}{(z_1-z_2)}+T^2(z_2)+...
\label{Tz1Tz2}
\end{equation}
Using the transformations of the primary field $\psi(\omega,\bar \omega)$ and conformal operators $L_{-2}^2(\omega), \bar L_{-2}^2(\bar \omega)$ under plane-cylinder mapping, the three point correlation function on the cylinder\\
$\langle \psi(\omega_1,\bar\omega_1)\varphi_1(\omega_2,\bar\omega_2)\psi(\omega_3,\bar\omega_3)\rangle$ can be written as
\begin{equation}
\langle \psi(\omega_1,\bar \omega_1)\varphi_1(\omega_2, \bar \omega_2)\psi(\omega_3,\bar \omega_3)\rangle=\left(\frac{2\pi}{L_x}\right)^{2\Delta+2\bar\Delta+4}(z_1z_3)^{\Delta}(\bar z_1 \bar z_3)^{\bar \Delta}\left(\Phi+\bar \Phi\right)
\label{threepointcyl1}
\end{equation}
where $\Phi$ is the three point correlation function on the plane and given by
\begin{equation}
\Phi=\langle \psi(z_1,\bar z_1)\left[z_2^4 L_{-2}^2(z_2)-\frac{c-10}{12}z_2^2 L_{-2}(z_2)+\frac{3 z_2^3}{2}L_{-3}(z_2)+\frac{\alpha c}{96}\right] \psi(z_3,\bar z_3)\rangle . \label{Phi}
\end{equation}
The correlation function $\langle \psi(z_1,\bar z_1)L_{-k}(z_2)\psi(z_3,\bar z_3)\rangle$ on the plane can be calculated with the help of
\begin{equation}
\langle \psi(z_1,\bar z_1)L_{-k}(z)\psi(z_2,\bar z_2)\rangle=\sum_{i=1}^2\left\{\frac{\Delta(k-1)}{(z_i-z)^k}-\frac{1}{(z_i-z)^{k-1}}\frac{\partial}{\partial z_i}\right\}\langle \psi(z_1,\bar z_1) \psi(z_2,\bar z_2)\rangle .
\label{threeL2}
\end{equation}
Now using expression for the two point correlation function on the plane given by Eq. (\ref{twopointplane}) one can easily write down the three point correlation functions $\Phi$ and $\bar \Phi$ on the plane. As result we obtain three point correlation function on the cylinder $\langle \psi(\omega_1,\bar\omega_1)\varphi_1(\omega_2,\bar\omega_2)\psi(\omega_3,\bar\omega_3)\rangle$ in the form
\begin{eqnarray}
\langle \psi(\omega_1,\bar \omega_1)\varphi_1(\omega_2, \bar \omega_2)\psi(\omega_3, \bar \omega_3)\rangle &=& \sum_{r=0}^{\infty}\sum_{\bar r=0}^{\infty} a_{r,r;\bar r, \bar r}(\xi_1\xi_2)^{\Delta+r} (\bar \xi_1 \bar\xi_2)^{\bar \Delta+\bar r}\nonumber\\
&+&\sum_{r_1\ne r_2} \sum_{\bar r_1 \ne \bar r_2} a_{r_1,r_2;\bar r_1, \bar r_2}\xi_1^{\Delta+r_1} \xi_2^{\Delta+r_2}\bar \xi_1^{\bar \Delta+\bar r_1}\bar \xi_2^{\bar \Delta+\bar r_2} , \label{three3}
\end{eqnarray}
where coefficient $a_{r,r;\bar r,\bar r}$ are given by
\begin{eqnarray}
&&a_{r,r;\bar r,\bar r}=\left(\frac{2\pi}{L_x}\right)^{2\Delta+2\bar\Delta+4}\nonumber\\&&c_r c_{\bar r}
\left\{\frac{\alpha c}{96}+(\Delta+r)\left[\Delta-\frac{c}{12}-\frac{1}{6}+\frac{r(5\Delta+1)(2\Delta+r)}{(\Delta+1)(2\Delta+1)}\right]+(r \to \bar r, \Delta \to \bar\Delta)\right\} .
\label{arrrr}
\end{eqnarray}
Now from Eqs. (\ref{matrixelem}),  (\ref{brr}) and (\ref{arrrr}) it is easy to obtain the universal structure constants $C_{n1n}$
\begin{eqnarray}
C_{n1n}=\left(\frac{L_x}{2\pi}\right)^4\frac{a_{r,r;\bar r,\bar r}}{b_r b_{\bar r}}&=&\frac{\alpha c}{96}+(\Delta+r)\left[\Delta-\frac{c}{12}-\frac{1}{6}+\frac{r(5\Delta+1)(2\Delta+r)}{(\Delta+1)(2\Delta+1)}\right]\nonumber\\
&+&(r \to \bar r, \Delta \to \bar\Delta)
\label{Cnkn}
\end{eqnarray}
which exactly reproduces Reinicke results for $\Delta \ne 0$ (see Eq. (2.17) and second line of Eq. (2.18) of \cite{reinicke87}).

For the case $\Delta=0$ one should replace the primary fields $\psi(\omega,\bar \omega)$ by tensor $T(\omega)\bar T(\bar \omega)$. The universal structure constants $C_{n1n}$ for the case $\Delta=0$ can be found along the same lines as above
\begin{equation}
C_{n1n}=\left[\left(\frac{c}{24}\right)^2+\frac{11c}{1440}+\left(\frac{11}{30}+\frac{c}{12}\right)r(2r^2-3)\right]+\left\{r \to \bar r\right\},
\quad r \ne 1, \bar r \ne 1 \label{cnknd0}
\end{equation}
which is exactly reproduce Reinicke results for $\Delta = 0$ (see Eq. (2.17) and first line of Eq. (2.18) of \cite{reinicke87}). Eqs. (\ref{Cnkn}) and (\ref{cnknd0}) can be combine in one equation which is given by Eq. (\ref{cnlnd0}).

\subsection{Universal structure constants for descendent states generated by the OPE of the primary fields.}
The asymptotic state $|h\rangle=\phi(0)|0\rangle$ created by a primary field $\phi(0)$ of conformal dimension $h$ is the source of an infinite tower of descendant states of higher conformal dimensions which can be obtained by inserting another primary field near 0 and applying operator product expansion (OPE). For such states the calculations of the universal structure constants $C_{n1n}$ is not straightforward. But for some cases it can be done. For example if the state under consideration is to be produced by the OPE $\psi(z) \phi(0)$ one can still use Eq. (\ref{matrixelem}), but instead of Eqs. (\ref{threepoint}) and (\ref{twopoint}) one should use the following equations
\begin{eqnarray}
\langle h|\psi(\omega_1,\bar \omega_1)\varphi_1(\omega_2, \bar \omega_2)\psi(\omega_3, \bar \omega_3)|h\rangle &=& \sum_{r_1,r_2;\bar r_1, \bar r_2} a_{r_1,r_2;\bar r_1, \bar r_2}\xi_1^{\Delta+r_1} \xi_2^{\Delta+r_2}\bar \xi_1^{\bar \Delta+\bar r_1}\bar \xi_2^{\bar \Delta+\bar r_2} \label{threepointnew} \\
\langle h|\psi(\omega_1,\bar \omega_1)\psi(\omega_3,\bar \omega_3)|h\rangle &=& \sum_{r,\bar r} b_{r,\bar r}\left(\xi_1 \xi_2\right)^{\Delta+r}\left(\bar \xi_1 \bar \xi_2\right)^{\bar \Delta+\bar r} \label{twopointnew}
\end{eqnarray}
where in $|h\rangle$ and out $\langle h|$ states can be defined as
\begin{eqnarray}
|h\rangle&=&\lim_{\zeta \to 0}\phi(\zeta)|0\rangle
\label{in}\\
\langle h|&=&\lim_{\zeta \to \infty}\zeta^{2h} \langle 0|\phi(\zeta).
\label{out}
\end{eqnarray}
The two point correlation functions on the cylinder can be written in terms of the two point correlation function on the infinite plane in the way similar to Eq. (\ref{twopointcyl})
\begin{eqnarray}
\langle h|\psi(\omega_1,\bar \omega_1)\psi(\omega_3,\bar \omega_3)|h\rangle&=&\left(\frac{2\pi}{L_x}\right)^{2\Delta+2\bar\Delta}\left(z_1 z_3\right)^{\Delta}\left(\bar z_1 \bar z_3\right)^{\bar \Delta} \langle h|\psi(z_1,\bar z_1)\psi(z_3,\bar z_3)|h\rangle
\label{twopointcyln}
\end{eqnarray}
Then using Eqs. (\ref{in}) and (\ref{out}) the two point correlation function on the infinite plane $\langle h|\psi(z_1,\bar z_1)\psi(z_3,\bar z_3)|h\rangle$  can be written as limiting case of four point correlation function $\langle \phi(\zeta_1,\bar \zeta_1)\psi(z_1,\bar z_1)\psi(z_3,\bar z_3) \phi(\zeta_3,\bar \zeta_3)\rangle$
\begin{equation}
\langle h|\psi(z_1,\bar z_1)\psi(z_3,\bar z_3)|h\rangle = \lim_{\zeta_1,\bar \zeta_1 \to \infty}\lim_{\zeta_3,\bar\zeta_3 \to 0}\zeta_1^{2h}{\bar \zeta_1}^{2 \bar h}\langle \phi(\zeta_1,\bar \zeta_1)\psi(z_1,\bar z_1)\psi(z_3,\bar z_3) \phi(\zeta_3,\bar \zeta_3)\rangle . \label{twotwo}
\end{equation}
Let us now calculate the three point correlation function on the cylinder \\
$\langle h| \psi(\omega_1,\bar \omega_1)\varphi_1(\omega_2, \bar \omega_2)\psi(\omega_3, \bar \omega_3)|h \rangle$. In the case when $\varphi_1(\omega_2, \bar \omega_2)= L_{-2}^2(\omega_2)+\bar L_{-2}^2(\bar \omega_2)$ one can consider holomorphic and antiholomorphic parts separately. The transformation of the conformal operator $L_{-2}^2(\omega)$  under plane-cylinder mapping is given by Eq. (\ref{L2transform}). Using the transformations of the primary field $\psi(\omega,\bar \omega)$ and conformal operators $L_{-2}^2(\omega), \bar L_{-2}^2(\bar \omega)$ under plane-cylinder mapping, the three point correlation function on the cylinder
$\langle h|\psi(\omega_1,\bar\omega_1)\varphi_1(\omega_2,\bar\omega_2)\psi(\omega_3,\bar\omega_3)|h \rangle$ can be written in terms of the three point correlation function on the infinite plane as
\begin{equation}
\langle h| \psi(\omega_1,\bar \omega_1)\varphi_1(\omega_2, \bar \omega_2)\psi(\omega_3,\bar \omega_3)|h \rangle=\left(\frac{2\pi}{L_x}\right)^{2\Delta+2\bar\Delta+4}\left(\Omega+\bar \Omega\right)
\label{threepointcylh1}
\end{equation}
where $\Omega$ is given by
\begin{eqnarray}
&&\Omega=(z_1z_3)^{\Delta}(\bar z_1 \bar z_3)^{\bar \Delta}\nonumber\\
&&\langle h| \psi(z_1,\bar z_1)\left[z_2^4 L_{-2}^2(z_2)-\frac{c-10}{12}z_2^2 L_{-2}(z_2)+\frac{3 z_2^3}{2}L_{-3}(z_2)+\frac{\alpha c}{96}\right]\psi(z_3,\bar z_3) |h \rangle \label{Xi}
\end{eqnarray}
The three point correlation function on the infinite  plane
 $\langle h|\psi(z_1,\bar z_1) \varphi_1(z_2, \bar z_2)\psi(z_3,\bar z_3)|h\rangle$ (with $\varphi_1 = L_{-2}^2, L_{-2}$ or $L_{-3}$) can be written as
\begin{eqnarray}
&&\langle h|\psi(z_1,\bar z_1) \varphi_1(z_2, \bar z_2)\psi(z_3,\bar z_3)|h\rangle = \nonumber\\
&&\lim_{\zeta_1 \to \infty}\lim_{\bar \zeta_1 \to \infty}
\zeta_1^{2h}{\bar \zeta_1}^{2 \bar h}\langle \phi(\zeta_1,\bar \zeta_1)\psi(z_1,\bar z_1) \varphi_1(z_2, \bar z_2)\psi(z_3,\bar z_3)\phi(0,0)\rangle . \label{threepointh}
\end{eqnarray}
Then the correlation function $\langle \phi(\zeta_1,\bar \zeta_1) \psi_{\Delta_1}(z_1,\bar z_1) L_{-k}(z_2) \psi_{\Delta_3}(z_3,\bar z_3)\phi(\zeta_3,\bar \zeta_3)\rangle$ can be calculated with the help of
\begin{eqnarray}
&&\langle  \psi_{1}(z_1) \psi_{2}(z_2)L_{-k}(z)\psi_{3}(z_3)\psi_{4}(z_4)\rangle=\nonumber\\&&\sum_{i=1}^4\left\{\frac{\Delta_i(k-1)}{(z_i-z)^k}
-\frac{1}{(z_i-z)^{k-1}}\frac{\partial}{\partial z_i}\right\} \langle \psi_{1}(z_1) \psi_2(z_2)\psi_{3}(z_3)\psi_{4}(z_4)\rangle
\label{threeL2d}
\end{eqnarray}
where $\Delta_i$ is conformal dimension of the primary field $\psi_i(z_i)$.

While the conformal invariance in general does not fix the precise form of the four-point correlation function, for some particular cases it is possible to write down an explicit form for it. For example in Ising model one can write the explicit form of the four-point correlation function $\langle \sigma(\zeta_1,\bar \zeta_1)\psi(z_1,\bar z_1)\psi(z_1,\bar z_3) \sigma(\zeta_3,\bar \zeta_3)\rangle$ as (see for example \cite{francesco} p. 446)
\begin{eqnarray}
&&\langle \sigma(\zeta_1,\bar \zeta_1)\psi(z_1,\bar z_1)\psi(z_1,\bar z_3) \sigma(\zeta_3,\bar \zeta_3)\rangle =\nonumber\\
&&\frac{\left(\sqrt{\frac{(z_1-\zeta_1)(z_3-\zeta_3)}{(z_1-\zeta_3)(z_3-\zeta_1)}}+\sqrt{\frac{(z_1-\zeta_3)
(z_3-\zeta_1)}{(z_1-\zeta_1)
(z_3-\zeta_3)}}\right)}{2(z_1-z_3)(\zeta_1-\zeta_3)^{1/8}} \frac{\left(\sqrt{\frac{(\bar z_1-\bar \zeta_1)(\bar z_3-\bar \zeta_3)}{(\bar z_1-\bar \zeta_3)(\bar z_3-\bar \zeta_1)}}+\sqrt{\frac{(\bar z_1-\bar \zeta_3)(\bar z_3-\bar \zeta_1)}{(\bar z_1-\bar \zeta_1)(\bar z_3-\bar \zeta_3)}}\right)}{2(\bar z_1-\bar z_3)(\bar \zeta_1-\bar \zeta_3)^{1/8}}.
\label{Isingfourpoint}
\end{eqnarray}
Here $\sigma(\zeta_i,\bar \zeta_i)$ is spin operator with conformal dimension $h=\bar h=1/16$ and $ \psi(z_i,\bar z_i)\equiv \psi(z_i)\bar \psi(\bar z_i)$ is the fermionic operator with conformal dimension $h=\bar h=1/2$. Then using Eqs. (\ref{twopointcyln}), (\ref{twotwo}) and (\ref{Isingfourpoint}) the two point correlation function on the cylinder $\langle h|\psi(\omega_1,\bar \omega_1)\psi(\omega_3,\bar \omega_3)|h\rangle$  can be written as
\begin{eqnarray}
&&\langle h|\psi(\omega_1,\bar \omega_1)\psi(\omega_3,\bar \omega_3)|h\rangle=\left(\frac{2\pi}{L_x}\right)^{2}\sqrt{z_1 z_3}\sqrt{\bar z_1 \bar z_3} \langle h|\psi(z_1,\bar z_1)\psi(z_3,\bar z_3)|h\rangle\nonumber\\
&&=\left(\frac{2\pi}{L_x}\right)^{2}\sqrt{z_1 z_3}\sqrt{\bar z_1 \bar z_3} \lim_{\zeta_1 \to \infty}\lim_{\bar \zeta_1 \to \infty}\zeta_1^{1/8}{\bar \zeta_1}^{1/8}\langle \sigma(\zeta_1,\bar \zeta_1)\psi(z_1,\bar z_1)\psi(z_1,\bar z_3) \sigma(0,0)\rangle \nonumber\\
&&=\left(\frac{2\pi}{L_x}\right)^{2}\frac{z_1+z_3}{2(z_3-z_1)}\frac{\bar z_1+\bar z_3}{2(\bar z_3-\bar z_1)}=\left(\frac{2\pi}{L_x}\right)^{2}\frac{1+\xi_1\xi_2}{2(1-\xi_1\xi_2)}\frac{1+\bar \xi_1 \bar \xi_2}{2(1-\bar \xi_1\bar \xi_2)}.\label{twotwoh}
\end{eqnarray}
Now the two point correlation function $\langle h|\psi(\omega_1,\bar \omega_1)\psi(\omega_3,\bar \omega_3)|h\rangle$ can be written in the form
\begin{equation}
 \langle h|\psi(\omega_1,\bar \omega_1)\psi(\omega_3,\bar \omega_3)|h\rangle=\sum_{r,\bar r} b_{r,\bar r}\left(\xi_1 \xi_2\right)^{r}\left(\bar \xi_1 \bar \xi_2\right)^{\bar r}
\label{U25h}
\end{equation}
where $b_{r,\bar r}$ is given by
\begin{equation}
b_{r,\bar r}=\left(\frac{2\pi}{L_x}\right)^2\left(1-\frac{1}{2}\delta_{r,0}\right)\left(1-\frac{1}{2}\delta_{\bar r,0}\right).
\label{brbarrh}
\end{equation}
The three point correlation function on the cylinder
$\langle h|\psi(\omega_1,\bar\omega_1)\varphi_1(\omega_2,\bar\omega_2)\psi(\omega_3,\bar\omega_3)|h \rangle$ for the Ising model can be written as
\begin{equation}
\langle h| \psi(\omega_1,\bar \omega_1)\varphi_1(\omega_2, \bar \omega_2)\psi(\omega_3,\bar \omega_3)|h \rangle=\left(\frac{2\pi}{L_x}\right)^{6}\left(\Omega+\bar \Omega\right)
\label{threepointcylh1I}
\end{equation}
where $\Omega$ is given by
\begin{equation}
\Omega=\sqrt{z_1 z_3}\sqrt{\bar z_1 \bar z_3}\langle h| \psi(z_1,\bar z_1)\left[z^4 L_{-2}^2(z)+\frac{19}{24}z^2 L_{-2}(z)+\frac{3 z^3}{2}L_{-3}(z)+\frac{49}{11520}\right]\psi(z_3,\bar z_3)|h \rangle .
\label{Xifin}
\end{equation}
Let us now calculate the three point correlation function \\ $z_2^2\sqrt{z_1 z_3}\sqrt{\bar z_1 \bar z_3}\mbox{$ \langle  \frac{1}{16}|$}\psi(z_1,\bar z_1) L_{-2}(z_2)\psi(z_3,\bar z_3)|\frac{1}{16} \rangle$ which we denote as $\langle L_{-2}\rangle$. From Eqs. (\ref{threepointh}), (\ref{threeL2d}) and (\ref{Isingfourpoint}) we can easily obtain that
\begin{eqnarray}
&&\langle L_{-2}\rangle=\frac{2z_2(z_2^2+z_1z_3)(z_1-3z_3)(z_3-3z_1)-z_2^2(z_1+z_3)^3-(z_1+z_3)(z_2^2+z_1 z_3)^2}{32(z_2-z_1)^2(z_2-z_3)^2(z_1-z_3)} \nonumber\\
&&\times \frac{\bar z_1+\bar z_3}{2(\bar z_3-\bar z_1)}\nonumber\\
&&=\frac{2(\xi_1+\xi_2)(1-3\xi_1\xi_2)(3-\xi_1\xi_2)+(1+\xi_1\xi_2)^3+(1+\xi_1\xi_2)(\xi_1+\xi_2)^2}{32(1-\xi_1)^2(1-\xi_2)^2(1-\xi_1 \xi_2)}\;\frac{1+\bar \xi_1 \bar \xi_3}{2(1-\bar \xi_1 \bar \xi_3)}. \nonumber\\ \label{threeL2h}
\end{eqnarray}
Now we can rewrite the Eq. (\ref{threeL2h}) keeping only the terms which contribute to the expansion of $(\xi_1 \xi_2)^r (\bar\xi_1\bar\xi_2)^{\bar r}$. As result we obtain
\begin{eqnarray}
\langle L_{-2}\rangle &\equiv& z_2^2\sqrt{z_1 z_3}\sqrt{\bar z_1 \bar z_3}\mbox{$ \langle  \frac{1}{16}|$}\psi(z_1,\bar z_1) L_{-2}(z_2)\psi(z_3,\bar z_3)\mbox{$|\frac{1}{16}\rangle$}\nonumber\\
&=&\frac{1+32\xi_1 \xi_2-\xi_1^2 \xi_2^2}{32(1-\xi_1 \xi_2)^2} \; \frac{1+\bar \xi_1 \bar \xi_2}{2(1-\bar \xi_1\bar \xi_2)}.\label{L2n}
\end{eqnarray}
Other three point functions can be found along the same line as above
\begin{eqnarray}
&&z_2^3\sqrt{z_1 z_3}\sqrt{\bar z_1 \bar z_3}\mbox{$ \langle  \frac{1}{16}|$}\psi(z_1,\bar z_1) L_{-3}(z_2)\psi(z_3,\bar z_3) \mbox{$\mbox{$|\frac{1}{16}\rangle$}$}=\nonumber\\
&&\frac{\xi_1^2 \xi_2^2-32\xi_1 \xi_2-1}{16(1-\xi_1 \xi_2)^2} \; \frac{1+\bar \xi_1 \bar \xi_2}{2(1-\bar \xi_1\bar \xi_2)} \label{L3n}\\
&&z_2^4\sqrt{z_1 z_3}\sqrt{\bar z_1 \bar z_3}\mbox{$ \langle  \frac{1}{16}|$}\psi(z_1,\bar z_1) L_{-2}^2(z_2)\psi(z_3,\bar z_3)\mbox{$\mbox{$|\frac{1}{16}\rangle$}$} \nonumber\\
&&=\frac{33+1662\xi_1 \xi_2+128\xi_1^2 \xi_2^2+1794\xi_1^3 \xi_2^3-33\xi_1^4 \xi_2^4}{512(1-\xi_1 \xi_2)^4}
\frac{1+\bar \xi_1 \bar \xi_2}{2(1-\bar \xi_1\bar \xi_2)}.\label{L22n}
\end{eqnarray}
Here again we keep only the terms which contribute to the expansion of $(\xi_1 \xi_2)^r (\bar\xi_1\bar\xi_2)^{\bar r}$. Now plugging Eqs. (\ref{twopointplane}), (\ref{L2n}) - (\ref{L22n}) back to Eqs. (\ref{Xifin}) we obtain that
\begin{eqnarray}
\Omega&=&\frac{7\left(\xi_1^4 \xi_2^4+478\xi_1^3 \xi_2^3+1920\xi_1^2 \xi_2^2+482\xi_1 \xi_2-1\right)}{2880(1-\xi_1 \xi_2)^4}\; \frac{1+\bar \xi_1 \bar \xi_2}{2(1-\bar \xi_1\bar \xi_2)}
\nonumber\\
&=&\left(\frac{7}{2880}-\frac{1687}{1440(1-\xi_1 \xi_2)}+\frac{49}{6(1-\xi_1 \xi_2)^2}-\frac{14}{(1-\xi_1 \xi_2)^3}+\frac{7}{(1-\xi_1 \xi_2)^4}\right)\; \frac{1+\bar \xi_1 \bar \xi_2}{2(1-\bar \xi_1\bar \xi_2)}.\nonumber\\
\label{threepointcylh1Ifin2}
\end{eqnarray}
Now  one can easy obtain
\begin{eqnarray}
\Omega &=&\sum_{r=0}^{\infty}\sum_{\bar r=0}^{\infty}\left(-\frac{7}{1440}+\frac{7 r^3}{6}\right)\left(1-\frac{1}{2}\delta_{r,0}\right)\left(1-\frac{1}{2}\delta_{\bar r,0}\right)(\xi_1 \xi_2)^{r}(\bar \xi_1 \bar \xi_2)^{\bar r}.\label{J4}
\end{eqnarray}
Thus three point correlation function on the cylinder
$\langle h|\psi(\omega_1,\bar\omega_1)\varphi_1(\omega_2,\bar\omega_2)\psi(\omega_3,\bar\omega_3)|h \rangle$ for the Ising model can be written as
\begin{eqnarray}
&&\langle h|\psi(\omega_1,\bar\omega_1)\varphi_1(\omega_2,\bar\omega_2)\psi(\omega_3,\bar\omega_3)|h \rangle=\nonumber\\
&&\sum_{r;\bar r} a_{r,r;\bar r, \bar r}(\xi_1 \xi_2)^{r}(\bar \xi_1 \bar \xi_2)^{\bar r}+\sum_{r_1 \ne r_2;\bar r_1 \ne \bar r_2} a_{r_1,r_2;\bar r_1, \bar r_2}\xi_1^{r_1} \xi_2^{r_2}\bar \xi_1^{\bar r_1}\bar \xi_2^{\bar r_2}
\label{3pointh}
\end{eqnarray}
where coefficient $a_{r,r;\bar r, \bar r}$ is given by
\begin{eqnarray}
a_{r,r;\bar r, \bar r}&=&\left(\frac{2\pi}{L_x}\right)^{6}\left(-\frac{7}{1440}+\frac{7 r^3}{6}\right)\left(1-\frac{1}{2}\delta_{r,0}\right)\left(1-\frac{1}{2}\delta_{\bar r,0}\right) +\left\{r \leftrightarrow \bar r\right\}.
\label{arrrr1h}
\end{eqnarray}
Now from Eqs. (\ref{brbarrh}) and (\ref{arrrr1h}) it is easy to obtain the universal structure constants $C_{n1n}$
\begin{eqnarray}
C_{n1n}=\left(\frac{L_x}{2\pi}\right)^4\frac{a_{r,r;\bar r,\bar r}}{b_r b_{\bar r}}=\left(-\frac{7}{1440}+\frac{7 r^3}{6}\right)
+(r \to \bar r).
\label{Cnknh}
\end{eqnarray}
Let us now consider the ground state $|0\rangle$ and the excited state $|L-p\rangle$
\begin{eqnarray}
|0\rangle&=&|\Delta_0=\frac{1}{16},r=0;\bar \Delta_0=0,\bar
r=0\rangle \label{Es0n}\\
|L-p\rangle&=&|\Delta_{L-p}=\frac{1}{16},r=p;\bar \Delta_{L-p}=0,\bar r=0\rangle . \label{EsLpn}
\end{eqnarray}
Since the $\Delta$ and $\bar \Delta$ is different in the ground state $|0\rangle$ and in the excited state $|L-p\rangle$ as well one has to treat the holomorphic and antiholomorphic dependence separately. Thus the universal structure constants $C_{010}$ for the ground state consist of a holomorphic part $C_{0}$ with $\Delta_0=\frac{1}{16}$ and $r=0$ and an antiholomorphic part $\bar C_0$ with $\bar \Delta_0=0$ and $\bar r=0$. The expression for $C_0$ can be obtained from the holomorphic part of the Eq. (\ref{cnlnd0}) and given by
\begin{equation}
C_0=-\frac{7}{1440}
\label{C0hol}
\end{equation}
and the expression for $\bar C_0$ can be obtained from the antiholomorphic part of the Eq. (\ref{cnlnd0}) and given by
\begin{equation}
\bar C_0=\frac{49}{11520}.
\label{C0antihol}
\end{equation}
Thus for the universal structure constants $C_{010}$ for the ground state we finally obtain
\begin{equation}
C_{010}=C_0+\bar C_0=-\frac{7}{1440}+\frac{49}{11520}=-\frac{7}{11520} . \label{C0k0}
\end{equation}
The universal structure constants $C_{(L-p)1(L-p)}$ for the exited state $|L-p \rangle$ consist of a holomorphic part $C_{L-p}$ with $\Delta=\frac{1}{16}$ and $r=p$ and an antiholomorphic part $\bar C_{L-p}$ with $\bar \Delta=0$ and $\bar r=0$. The expression for $C_{L-p}$ can be obtained from the holomorphic part of the  Eq. (\ref{Cnknh}) and given by
\begin{equation}
C_{L-p}=-\frac{7}{1440}+\frac{7 p^3}{6}
\label{CLphol}
\end{equation}
and expression for $\bar C_{L-p}$ can be obtained from the antiholomorphic part of the Eq. (\ref{cnlnd0}) and given by
\begin{equation}
\bar C_{L-p}=\frac{49}{11520}.
\label{CLpantihol}
\end{equation}
Thus for the universal structure constants $C_{(L-p)1(L-p)}$ for the exited state $|L-p \rangle$ we finally obtain
\begin{equation}
C_{(L-p)1(L-p)}=C_{L-p}+\bar C_{L-p}=-\frac{7}{11520}+\frac{7 p^3}{6}.\label{Cnknn}
\end{equation}
Now we can calculate the ratio $r_{L-p}$ which is given by
\begin{eqnarray}
r_{L-p}&=&\frac{C_{(L-p)1(L-p)}-C_{010}}{C_{010}}=-1920 p^3 . \label{rLpconfn}
\end{eqnarray}
Thus, we can see that  the ratio $r_{L-p}$ given by Eqs. (\ref{rLpconfn}) coincide with Eq. (\ref{rL2}) for all p. Thus we verify the universality of the aspect ratios $r_{L-p}$ for all values of $p$.

\section{Conclusion}
\label{concl}
For the Ising model on an infinitely long cylinder there are six different boundary universality classes with different values of $\Delta$ and $\bar \Delta$. Past efforts have been focused mainly on periodic ($\Delta = 0, \bar \Delta = 0$) and antiperiodic boundary conditions ($\Delta = 1/16, \bar \Delta = 1/16$). Little attention has been paid to the boundary condition given by Eqs. (\ref{eq7.3}) - (\ref{eq7.6}). In this paper we fill partially this gap and consider one of those boundary conditions, namely those given by Eq. (\ref{eq7.4}) with $(\Delta, \bar \Delta)= \left(\frac{1}{16}, 0\right)$. These boundary conditions, which we call duality-twisted,  may be interpreted as inserting a specific defect line (``seam'') in the system, along non contractible circles of the cylinder, before closing it into a torus. In this work we derive exact expressions for ${\it all}$ eigenvalues of the transfer matrix for the critical ferromagnetic Ising model on the $M \times N$ square lattice wrap on torus with specific defect line (``seam'') with seam $(r,s)=(1,2)$. We reproduce by an exact calculation the exact formula for the universal conformal partition function $Z_{(1,2)}(q)$, for which there is also an anterior numerical confirmation \cite{Pearce2001}.

In the limit $N \to \infty$ we obtain the asymptotic expansion of the free energy $f$ and the inverse correlation lengths $\xi_p^{-1}$ and $\xi_{L-p}^{-1}$ for an infinitely long cylinder of circumference $L_x$ with duality twisted boundary conditions. We find that subdominant finite-size corrections to scaling should be to the form $a_k/L^{2k-1}_x$ for the free energy $f$ and $b_k(p)/L_x^{2k-1}$ $c_k(p)/L_x^{2k-1}$ and for inverse correlation lengths $\xi^{-1}_p$ and $\xi^{-1}_{L-p}$, respectively, with integer value of $k$. We investigate the sets $\left\{a_k, b_k(p), c_k(p)\right\}$ by exact evaluation and find that the amplitude ratios $b_k(p)/a_k$ and $c_k(p)/a_k$ are universal.  We verify this universal behavior in the framework of perturbating conformal approach by using the Reinicke formula \cite{reinicke87} for descendent states ($L_{-1}^n|\Delta\rangle$) of the primary field $\psi(z)$ with conformal dimension $\Delta = 1/2$ given by Eq. (\ref{cnlnd0}) and also using our computation of the universal structure constant $C_{n1n}$ for descendent states generated by the OPE of the primary fields $\psi(z)\sigma(0)$.

\section{Acknowledgement}
 This research was partially supported by the grant of the Science Committee of the Ministry of Science and Education of the Republic of Armenia under contract 15T-1C068. N.I. and R.K. was also supported by IRSES grants (DIONICOS, PIRSES-GA-2013-612707) within the 7th EU Framework Programme. We would like to thank Paul Pearce and Rubik Poghossian for valuable discussions and comments.


\appendix
\section{The Euler-Maclaurin and Boole summation formulas}
Let us start with the Euler - Maclaurin summation formula.
Suppose that $f(x)$ together with its derivatives is continuous within the interval $(0, b)$. The Euler-Maclaurin summation formula states
\begin{eqnarray}
\label{general EM}
\sum_{k=0}^{L-1}f(k h+\alpha h)=\frac{1}{h}\int_0^b f(x)dx+\sum_{n=1}^{\infty}\frac{h^{n-1}}{n!}
B_{n}(\alpha)(f^{(n-1)}(b)-f^{(n-1)}(0))
\end{eqnarray}
where $0\leq\alpha\leq1$, $h=\frac{b}{L}$ and $B_n(\alpha)$ are the Bernoulli polynomials. The Bernoulli polynomials are characterized by a generating function
\begin{eqnarray}
\frac{t e^{x t}}{e^t-1}=\sum_{n=0}^{\infty}B_n(x)\frac{t^n}{n!}.
\end{eqnarray}
For example the first few Bernoulli polynomials are
\begin{eqnarray}
B_0(\alpha)&=&1\nonumber\\
B_1(\alpha)&=&\alpha-\frac{1}{2}\nonumber\\
B_2(\alpha)&=&\alpha^2-\alpha+\frac{1}{6}\nonumber\\
B_3(\alpha)&=&\alpha^3-\frac{3\alpha^2}{2}+\frac{\alpha}{2}\nonumber\\
B_4(\alpha)&=&\alpha^4-2\alpha^3+\alpha^2-\frac{1}{30},\nonumber
\end{eqnarray}
$B_p \equiv B_p(0)$ are the Bernoulli numbers.
Starting from $p=3$ the values of $B_p$ with odd $p$ are zero. For $\alpha=\frac{1}{2}$ we have
\begin{eqnarray}
\label{B 0 1/2}
&&B_n(1/2)=(2^{1-n}-1)B_n
\end{eqnarray}
For alternating finite sums one can use the Boole summation formula \cite{Borwein}
\begin{eqnarray}
\label{BooleO}
\sum_{k=0}^{L-1}(-1)^k f(k h+\alpha h)=\frac{1}{2}\sum_{n=0}^{\infty}\frac{h^{n}}{n!}
E_{n}(\alpha)\left[(-1)^{L-1} f^{(n)}(b)+f^{(n)}(0)\right]
\end{eqnarray}
where $0\leq\alpha\leq1$, $h=\frac{b}{L}$ and $E_n(\alpha)$ are the Euler polynomials. The Euler polynomials are characterized by a generating function
\begin{eqnarray}
\frac{2 e^{x t}}{e^t+1}=\sum_{n=0}^{\infty}E_n(x)\frac{t^n}{n!}.
\end{eqnarray}
For example the first few Euler polynomials are
\begin{eqnarray}
E_0(\alpha)&=&1\nonumber\\
E_1(\alpha)&=&\alpha-\frac{1}{2}\nonumber\\
E_2(\alpha)&=&\alpha^2-\alpha\nonumber\\
E_3(\alpha)&=&\alpha^3-\frac{3\alpha^2}{2}+\frac{1}{4}\nonumber\\
E_4(\alpha)&=&\alpha^4-2\alpha^3+\alpha\nonumber .
\end{eqnarray}

All even Euler polynomials at $\alpha=0$ are zero ($E_{2p}(0)=0$). Thus for $\alpha=0$ the Boole summation formula now reads as

\begin{equation}
\sum_{k=0}^{L-1}(-1)^k f(k h)=\frac{1}{2}\sum_{n=0}^{\infty}\frac{h^{2n+1}}{(2n+1)!}
E_{2n+1}(0)\left[(-1)^{L-1} f^{(2n+1)}(b)+f^{(2n+1)}(0)\right].\label{Boole}
\end{equation}

\end{document}